\documentclass[copyright,creativecommons]{eptcs}
\providecommand{\event}{QPL 2021} 

\usepackage{amsmath,amssymb,amsthm,mathtools}
\usepackage{marvosym}
\usepackage{tikz}
\usetikzlibrary{arrows.meta}
\usetikzlibrary{decorations.pathmorphing,shapes}
\usetikzlibrary{circuits.ee.IEC}
\usepackage{dsfont}
\usepackage{soul}
\usepackage[normalem]{ulem}
\usepackage{enumerate}
\usepackage{subfig} 
\usepackage{wrapfig}
\usepackage{footmisc}

\usepackage{tocbasic}
\DeclareTOCStyleEntry[
  beforeskip=.2em plus 1pt,
  pagenumberformat=\textbf
]{tocline}{section}

\theoremstyle{plain}
\newtheorem{theorem}[equation]{Theorem}
\newtheorem{lemma}[equation]{Lemma}
\newtheorem{proposition}[equation]{Proposition}
\newtheorem{corollary}[equation]{Corollary}
\theoremstyle{definition}
\newtheorem{definition}[equation]{Definition}
\newtheorem{construction}[equation]{Construction}

\newtheorem{question}[equation]{Question}
\newtheorem{problem}[equation]{Problem}
\newtheorem{example}[equation]{Example}
\newtheorem{exercise}[equation]{Exercise}
\newtheorem*{answer}{Answer}
\newtheorem*{solution}{Solution}
\newtheorem{remark}[equation]{Remark}

\newtheorem{notation}[equation]{Notation}

\newcommand\define[1]{\emph{\textbf{#1}}}

\numberwithin{equation}{section}

 \let\b=\beta  \let\de=\delta 
  \let\q=\theta  
\let\l=\lambda 
\let\s=\sigma    \let\c=\chi 
       \let\D=\Delta  
     
\let\C=\Chi

\newcommand{\be}{\begin{equation}}
\newcommand{\ee}{\end{equation}}
\def\ba{\begin{align}} 
\def\ea{\end{align}}
\newcommand{\bea}{\begin{eqnarray}}
\newcommand{\eea}{\end{eqnarray}}
\newcommand{\bx}{\begin{example}}
\newcommand{\ex}{\end{example}}
\newcommand{\bex}{\begin{exercise}}
\newcommand{\eex}{\end{exercise}}
\newcommand{\ban}{\begin{answer}}
\newcommand{\ean}{\end{answer}}
\newcommand{\bt}{\begin{theorem}}
\newcommand{\et}{\end{theorem}}
\newcommand{\bc}{\begin{corollary}}
\newcommand{\ec}{\end{corollary}}
\newcommand{\blem}{\begin{lemma}}
\newcommand{\elem}{\end{lemma}}
\newcommand{\bp}{\begin{problem}}
\newcommand{\ep}{\end{problem}}
\newcommand{\bn}{\begin{proposition}}
\newcommand{\en}{\end{proposition}}
\newcommand{\bd}{\begin{definition}}
\newcommand{\ed}{\end{definition}}
\newcommand{\bcon}{\begin{construction}}
\newcommand{\econ}{\end{construction}}
\newcommand{\bq}{\begin{question}}
\newcommand{\eq}{\end{question}}
\newcommand{\bprf}{\begin{proof}}
\newcommand{\eprf}{\end{proof}}
\newcommand{\br}{\begin{remark}}
\newcommand{\er}{\end{remark}}
\newcommand{\bs}{\begin{solution}}
\newcommand{\es}{\end{solution}}
\newcommand{\beqs}{\begin{eqnarray}}
\newcommand{\eeqs}{\end{eqnarray}}

\let\p=\partial \let\ov=\overline

\newcommand{\<}{\langle}
\renewcommand{\>}{\rangle}

\newcommand{\id}{\mathrm{id}}

\newcommand{\mC}{\mathcal{C}}
\newcommand{\mM}{\mathcal{M}}

\newcommand{\up}{\uparrow}
\newcommand{\dn}{\downarrow}

\newcommand{\Lra}{\Leftrightarrow}

\newcommand{\tr}{{\rm tr} }

\def\R{{{\mathbb R}}}
\def\C{{{\mathbb C}}}

\def\N{{{\mathbb N}}}
\def\Z{{{\mathbb Z}}}

\newcommand{\Ad}{\mathrm{Ad}}

\newcommand{\fdCAlgPU}{\mathbf{fdC\text{*-}AlgPU}}

\newcommand{\fdCAlgUY}{\mathbf{fdC\text{*-}AlgU}_{\text{\Yinyang}}}

\def\mA{{{\mathcal{A}}}}

\def\mB{{{\mathcal{B}}}}

\newcommand{\op}{\mathrm{op}}

\newcommand{\FinStoch}{\mathbf{FinStoch}}

\makeatletter
\renewcommand*\env@matrix[1][\arraystretch]{%
  \edef\arraystretch{#1}%
  \hskip -\arraycolsep
  \let\@ifnextchar\new@ifnextchar
  \array{*\c@MaxMatrixCols c}}
\makeatother

\newcounter{sarrow}
\newcommand\xstoch[1]{%
\stepcounter{sarrow}%
\mathrel{\begin{tikzpicture}[baseline= {( $ (current bounding box.south) + (0,-0.1ex) $ )}]
\node[inner sep=.5ex] (\thesarrow) {\;$\scriptstyle #1$\;};
\path[draw,{<[scale=1.5,width=3,length=2]}-,decorate,
  decoration={snake,amplitude=0.3mm,segment length=2.1mm,pre=lineto,pre length=1pt}] 
    (\thesarrow.south east) -- (\thesarrow.south west);
\end{tikzpicture}}%
}

\newlength\stateheight
\newlength\minimumstatewidth
\tikzset{width/.initial=\minimummorphismwidth}
\tikzset{colour/.initial=white}

\newif\ifblack\pgfkeys{/tikz/black/.is if=black}
\newif\ifwedge\pgfkeys{/tikz/wedge/.is if=wedge}
\newif\ifvflip\pgfkeys{/tikz/vflip/.is if=vflip}
\newif\ifhflip\pgfkeys{/tikz/hflip/.is if=hflip}
\newif\ifhvflip\pgfkeys{/tikz/hvflip/.is if=hvflip}

\pgfdeclarelayer{foreground}
\pgfdeclarelayer{background}
\pgfsetlayers{background,main,foreground}

\def\thickness{0.4pt}

\makeatletter
\pgfkeys{%
  /tikz/on layer/.code={
    \pgfonlayer{#1}\begingroup
    \aftergroup\endpgfonlayer
    \aftergroup\endgroup
  },
  /tikz/node on layer/.code={
    \gdef\node@@on@layer{%
      \setbox\tikz@tempbox=\hbox\bgroup\pgfonlayer{#1}\unhbox\tikz@tempbox\endpgfonlayer\pgfsetlinewidth{\thickness}\egroup}
    \aftergroup\node@on@layer
  },
  /tikz/end node on layer/.code={
    \endpgfonlayer\endgroup\endgroup
  }
}
\def\node@on@layer{\aftergroup\node@@on@layer}
\makeatother

\makeatletter
\pgfdeclareshape{state}
{
    \savedanchor\centerpoint
    {
        \pgf@x=0pt
        \pgf@y=0pt
    }
    \anchor{center}{\centerpoint}
    \anchorborder{\centerpoint}
    \saveddimen\overallwidth
    {
        \pgf@x=3\wd\pgfnodeparttextbox
        \ifdim\pgf@x<\minimumstatewidth
            \pgf@x=\minimumstatewidth
        \fi
    }
    \savedanchor{\upperrightcorner}
    {
        \pgf@x=.5\wd\pgfnodeparttextbox
        \pgf@y=.5\ht\pgfnodeparttextbox
        \advance\pgf@y by -.5\dp\pgfnodeparttextbox
    }
    \anchor{A}
    {
        \pgf@x=-\overallwidth
        \divide\pgf@x by 4
        \pgf@y=0pt
    }
    \anchor{B}
    {
        \pgf@x=\overallwidth
        \divide\pgf@x by 4
        \pgf@y=0pt
    }
    \anchor{text}
    {
        \upperrightcorner
        \pgf@x=-\pgf@x
        \ifhflip
            \pgf@y=-\pgf@y
            \advance\pgf@y by 0.4\stateheight
        \else
            \pgf@y=-\pgf@y
            \advance\pgf@y by -0.4\stateheight
        \fi
    }
    \backgroundpath
    {
       \begin{pgfonlayer}{foreground}
        \pgfsetstrokecolor{black}
        \pgfsetlinewidth{\thickness}
        \pgfpathmoveto{\pgfpoint{-0.5*\overallwidth}{0}}
        \pgfpathlineto{\pgfpoint{0.5*\overallwidth}{0}}
        \ifhflip
            \pgfpathlineto{\pgfpoint{0}{\stateheight}}
        \else
            \pgfpathlineto{\pgfpoint{0}{-\stateheight}}
        \fi
        \pgfpathclose
        \ifblack
            \pgfsetfillcolor{black!50}
            \pgfusepath{fill,stroke}
        \else
            \pgfusepath{stroke}
        \fi
       \end{pgfonlayer}
    }
}

\makeatletter
\pgfdeclareshape{effect}
{
    \savedanchor\centerpoint
    {
        \pgf@x=0pt
        \pgf@y=0pt
    }
    \anchor{center}{\centerpoint}
    \anchorborder{\centerpoint}
    \saveddimen\overallwidth
    {
        \pgf@x=3\wd\pgfnodeparttextbox
        \ifdim\pgf@x<\minimumstatewidth
            \pgf@x=\minimumstatewidth
        \fi
    }
    \savedanchor{\upperrightcorner}
    {
        \pgf@x=.5\wd\pgfnodeparttextbox
        \pgf@y=-.5\ht\pgfnodeparttextbox 
        \advance\pgf@y by .5\dp\pgfnodeparttextbox 
    }
    \anchor{A}
    {
        \pgf@x=-\overallwidth
        \divide\pgf@x by 4
        \pgf@y=0pt
    }
    \anchor{B}
    {
        \pgf@x=\overallwidth
        \divide\pgf@x by 4
        \pgf@y=0pt
    }
    \anchor{text}
    {
        \upperrightcorner
        \pgf@x=-\pgf@x
        \ifhflip
            \pgf@y=\pgf@y 
            \advance\pgf@y by -0.4\stateheight 
        \else
            \pgf@y=\pgf@y 
            \advance\pgf@y by 0.4\stateheight 
        \fi
    }
    \backgroundpath
    {
       \begin{pgfonlayer}{foreground}
        \pgfsetstrokecolor{black}
        \pgfsetlinewidth{\thickness}
        \pgfpathmoveto{\pgfpoint{-0.5*\overallwidth}{0}}
        \pgfpathlineto{\pgfpoint{0.5*\overallwidth}{0}}
        \ifhflip
            \pgfpathlineto{\pgfpoint{0}{-\stateheight}}
        \else
            \pgfpathlineto{\pgfpoint{0}{\stateheight}}
        \fi
        \pgfpathclose
        \ifblack
            \pgfsetfillcolor{black!50}
            \pgfusepath{fill,stroke}
        \else
            \pgfusepath{stroke}
        \fi
       \end{pgfonlayer}
    }
}

\pgfdeclareshape{my ground}
{
  \inheritsavedanchors[from=rectangle ee]
  \inheritanchor[from=rectangle ee]{center}
  \inheritanchor[from=rectangle ee]{north}
  \inheritanchor[from=rectangle ee]{south}
  \inheritanchor[from=rectangle ee]{east}
  \inheritanchor[from=rectangle ee]{west}
  \inheritanchor[from=rectangle ee]{north east}
  \inheritanchor[from=rectangle ee]{north west}
  \inheritanchor[from=rectangle ee]{south east}
  \inheritanchor[from=rectangle ee]{south west}
  \inheritanchor[from=rectangle ee]{input}
  \inheritanchor[from=rectangle ee]{output}
  \anchorborder{\csname pgf@anchor@my ground@input\endcsname}

  \backgroundpath{
    \pgf@process{\pgfpointadd{\southwest}{\pgfpoint{\pgfkeysvalueof{/pgf/outer xsep}}{\pgfkeysvalueof{/pgf/outer ysep}}}}
    \pgf@xa=\pgf@x \pgf@ya=\pgf@y
    \pgf@process{\pgfpointadd{\northeast}{\pgfpointscale{-1}{\pgfpoint{\pgfkeysvalueof{/pgf/outer xsep}}{\pgfkeysvalueof{/pgf/outer ysep}}}}}
    \pgf@xb=\pgf@x \pgf@yb=\pgf@y
    \pgfpathmoveto{\pgfqpoint{\pgf@xa}{\pgf@ya}}
    \pgfpathlineto{\pgfqpoint{\pgf@xa}{\pgf@yb}}
    \pgfmathsetlength\pgf@xa{.5\pgf@xa+.5\pgf@xb}
    \pgfmathsetlength\pgf@yc{.16666\pgf@yb-.16666\pgf@ya}
    \advance\pgf@ya by\pgf@yc
    \advance\pgf@yb by-\pgf@yc
    \pgfpathmoveto{\pgfqpoint{\pgf@xa}{\pgf@ya}}
    \pgfpathlineto{\pgfqpoint{\pgf@xa}{\pgf@yb}}
    \advance\pgf@ya by\pgf@yc
    \advance\pgf@yb by-\pgf@yc
    \pgfpathmoveto{\pgfqpoint{\pgf@xb}{\pgf@ya}}
    \pgfpathlineto{\pgfqpoint{\pgf@xb}{\pgf@yb}}
  }
}
\makeatother

\tikzset{inline text/.style =
  {text height=1.2ex,text depth=0.25ex,yshift=0.5mm}}
\tikzset{arrow box/.style =
  {rectangle,inline text,fill=white,draw,
    minimum height=5mm,yshift=-0.5mm,minimum width=5mm}}
\tikzset{bubble/.style =
  {inner sep=0mm,minimum width=3mm,minimum height=3mm,
    draw,shape=circle,fill=white}}
\tikzset{dot/.style =
  {inner sep=0mm,minimum width=1mm,minimum height=1mm,
    draw,shape=circle}}
\tikzset{white dot/.style = {dot,fill=white,text depth=-0.2mm}}
\tikzset{scalar/.style = {diamond,draw,inner sep=1pt}}
\tikzset{square/.style =
  {inner sep=0mm,minimum width=2mm,minimum height=2mm,
    draw,shape=rectangle}}

\tikzset{star/.style = {dot,fill=white,text depth=-0.2mm}}
\tikzset{copier/.style = {dot,fill,text depth=-0.2mm}}
\tikzset{fakecopier/.style = {square,fill,text depth=-0.2mm}}
\tikzset{discarder/.style = {my ground,draw,inner sep=0pt,
    minimum width=4.2pt,minimum height=11.2pt,anchor=input,rotate=90}}

\tikzset{xshiftu/.style = {shift = {(#1, 0)}}}
\tikzset{yshiftu/.style = {shift = {(0, #1)}}}
\tikzset{scriptstyle/.style={font=\everymath\expandafter{\the\everymath\scriptstyle}}}

\title{Conditional Distributions for Quantum Systems}
\author{Arthur J.\ Parzygnat
\institute{Institut des Hautes \'Etudes Scientifiques\\ Bures-sur-Yvette, France}
\email{parzygnat@ihes.fr}
}
\def\titlerunning{Conditional Distributions}
\def\authorrunning{A.J.\ Parzygnat}
\begin{document}
\maketitle

\begin{abstract}
Conditional distributions, as defined by the Markov category framework, are studied in the setting of matrix algebras (quantum systems). Their construction as linear unital maps are obtained via a categorical Bayesian inversion procedure. Simple criteria establishing when such linear maps are positive are obtained. Several examples are provided, including the standard EPR scenario, where the EPR correlations are reproduced in a purely compositional (categorical) manner. A comparison between the Bayes map, the Petz recovery map, and the Leifer--Spekkens acausal belief propagation is provided, illustrating some similarities and key differences. 

\end{abstract}

\tableofcontents

\section{Introduction}

There is a correspondence between stochastic maps (conditional probabilities) on finite sets and positive unital maps on finite-dimensional commutative $C^*$-algebras by a stochastic variant of Gelfand duality~\cite{Pa17,FuJa15}. Hence, any concept involving stochastic maps that is described categorically can be transferred to arbitrary (not necessarily commutative) $C^*$-algebras, thus justifying quantum analogues of classical concepts. In particular, Bayesian inversion, disintegrations, and conditioning have been formulated categorically~\cite{Fo12,CuSt14,DDGK16,CDDG17,ChJa18,Ja18EPTCS,PaRu19,Fr19}, and the first two have been analyzed in the setting of finite-dimensional $C^*$-algebras in~\cite{PaRu19,PaRuBayes} through a generalization of Markov categories~\cite{ChJa18,Fr19} to their quantum variants~\cite{PaBayes}. However, conditioning in this setting remains unexplored, as far as I am aware.%
\footnote{The conditioning in~\cite{Ja18EPTCS} does not use the multiplication map in its formulation of conditioning in the quantum setting.}
Since conditioning in quantum theory is not a settled subject, this direction deserves some investigation. 

The purpose of the present work is to begin the systematic study of quantum conditionals as positive maps between finite-dimensional $C^*$-algebras. Although this goal is not fully realized here, we are content with achieving it on bi-partite systems of matrix algebras equipped with states whose marginals are faithful. Even though this sounds quite restrictive, it already includes many cases of interest, such as the fully entangled EPR state on a two qubit system~\cite{EPR,Bo51}. The case of multi-partite states, non-faithful marginals, and more general hybrid classical-quantum systems will be addressed in future work. 

In this work, we use category theory to \emph{define} what we mean by quantum conditionals. Then, we prove a purely categorical theorem indicating how one can \emph{construct} quantum conditionals through the usage of \emph{Bayes maps} (whose definition is motivated by categorical probability theory). We then implement this construction in the setting of matrix algebras. In general, the resulting conditional does not define a positive map. As such, we find 
necessary and sufficient conditions for conditionals to be positive. A positive conditional need not be \emph{completely} positive, and EPR provides an example illustrating this point. We end by introducing the \emph{conditional domain}, which is the largest operator system for which a conditional is positive (in the Heisenberg picture). Typically, this operator system is not a $C^*$-subalgebra. Examples are provided throughout. 

\section{Quantum Markov categories}

This section briefly reviews the abstract theory of quantum CD and Markov categories~\cite{PaBayes}, which are generalizations of CD and Markov categories~\cite{ChJa18,Fr19}. String diagrams are reviewed in these mentioned papers, but see~\cite{Se10} for a more thorough exposition. Time will always go up the page. The composition will go up the page for definitions and the example $\FinStoch$, while the composition will go down the page for $C^*$-algebra maps (in the Heisenberg picture). 

\bd
A \define{classical CD category} is a symmetric monoidal category $(\mM,\otimes,I)$, with $\otimes$ the tensor product and $I$ the unit (associators and unitors are excluded from the notation), and where each object $X$ in $\mM$ is equipped with morphisms $!_{X}\equiv\vcenter{\hbox{%
\begin{tikzpicture}[font=\footnotesize]
\node[discarder] (d) at (0,0.15) {};
\node at (0.15,-0.15) {\scriptsize $X$};
\draw (d) to (0,-0.25);
\end{tikzpicture}}}: X\to I$, called the \define{discarder/grounding}, and $\Delta_{X}\equiv\vcenter{\hbox{%
\begin{tikzpicture}
\node[copier] (c) at (0,0.4) {};
\draw (c)
to[out=15,in=-90] (0.25,0.7);
\draw (c)
to[out=165,in=-90] (-0.25,0.7);
\draw (c) to (0,0.1);
\end{tikzpicture}}}: X\to X\otimes X,$ called the \define{copy/duplicate}, all satisfying the following conditions 
\be
\label{eq:MarkovCatA}
\vcenter{\hbox{%
\begin{tikzpicture}[font=\footnotesize]
\node[copier] (c) at (0,0) {};
\coordinate (x1) at (-0.3,0.3);
\node[discarder] (d) at (x1) {};
\coordinate (x2) at (0.3,0.5);
\draw (c) to[out=165,in=-90] (x1);
\draw (c) to[out=15,in=-90] (x2);
\draw (c) to (0,-0.3);
\end{tikzpicture}}}
\quad=\quad
\vcenter{\hbox{%
\begin{tikzpicture}[font=\footnotesize]
\draw (0,0) to (0,0.8);
\end{tikzpicture}}}
\quad=\quad
\vcenter{\hbox{%
\begin{tikzpicture}[font=\footnotesize]
\node[copier] (c) at (0,0) {};
\coordinate (x1) at (-0.3,0.5);
\coordinate (x2) at (0.3,0.3);
\node[discarder] (d) at (x2) {};
\draw (c) to[out=165,in=-90] (x1);
\draw (c) to[out=15,in=-90] (x2);
\draw (c) to (0,-0.3);
\end{tikzpicture}}}
\qquad\qquad
\vcenter{\hbox{%
\begin{tikzpicture}[font=\footnotesize]
\node[copier] (c2) at (0,0) {};
\node[copier] (c1) at (-0.3,0.3) {};
\draw (c2) to[out=165,in=-90] (c1);
\draw (c2) to[out=15,in=-90] (0.4,0.6);
\draw (c1) to[out=165,in=-90] (-0.6,0.6);
\draw (c1) to[out=15,in=-90] (0,0.6);
\draw (c2) to (0,-0.3);
\end{tikzpicture}}}
\quad=\quad
\vcenter{\hbox{%
\begin{tikzpicture}[font=\footnotesize]
\node[copier] (c2) at (-0.5,0) {};
\node[copier] (c1) at (-0.2,0.3) {};
\draw (c2) to[out=15,in=-90] (c1);
\draw (c2) to[out=165,in=-90] (-1,0.6);
\draw (c1) to[out=165,in=-90] (-0.5,0.6);
\draw (c1) to[out=15,in=-90] (0.1,0.6);
\draw (c2) to (-0.5,-0.3);
\end{tikzpicture}}}
\qquad\qquad
\vcenter{\hbox{%
\begin{tikzpicture}[font=\small]
\node[copier] (c) at (0,0.4) {};
\draw (c)
to[out=15,in=-90] (0.25,0.65)
to[out=90,in=-90] (-0.25,1.2);
\draw (c)
to[out=165,in=-90] (-0.25,0.65)
to[out=90,in=-90] (0.25,1.2);
\draw (c) to (0,0.1);
\end{tikzpicture}}}
\quad=\quad
\vcenter{\hbox{%
\begin{tikzpicture}[font=\small]
\node[copier] (c) at (0,0.4) {};
\draw (c)
to[out=15,in=-90] (0.25,0.7);
\draw (c)
to[out=165,in=-90] (-0.25,0.7);
\draw (c) to (0,0.1);
\end{tikzpicture}}}
\ee
\be
\label{eq:MarkovCatB}
\vcenter{\hbox{%
\begin{tikzpicture}[font=\footnotesize]
\node[discarder] (d) at (0,0) {};
\draw (d) to +(0,-0.5);
\node at (0.5,-0.3) {$X\otimes Y$};
\end{tikzpicture}}}
=
\vcenter{\hbox{%
\begin{tikzpicture}[font=\footnotesize]
\node[discarder] (d) at (0,0) {};
\node[discarder] (d2) at (0.6,0) {};
\draw (d) to +(0,-0.5);
\draw (d2) to +(0,-0.5);
\node at (0.2,-0.3) {$X$};
\node at (0.8,-0.3) {$Y$};
\end{tikzpicture}}}
\qquad\quad
\vcenter{\hbox{%
\begin{tikzpicture}[font=\footnotesize]
\node[discarder] (d) at (0,0) {};
\draw (d) to +(0,-0.5);
\node at (0.2,-0.3) {$I$};
\end{tikzpicture}}}
=\;
\vcenter{\hbox{%
\begin{tikzpicture}[font=\footnotesize]
\node at (0.2,-0.05) {};
\draw [gray,dashed] (0,0) rectangle (0.45,0.65);
\end{tikzpicture}}}
\qquad\quad
\vcenter{\hbox{%
\begin{tikzpicture}[font=\footnotesize]
\node[copier] (c) at (0,0.4) {};
\draw (c)
to[out=15,in=-90] (0.25,0.7);
\draw (c)
to[out=165,in=-90] (-0.25,0.7);
\draw (c) to (0,0);
\node at (0.5,0.1) {$X\otimes Y$};
\end{tikzpicture}}}
=
\vcenter{\hbox{%
\begin{tikzpicture}[font=\footnotesize]
\node[copier] (c) at (0,0) {};
\node[copier] (c2) at (0.4,0) {};
\draw (c) to[out=15,in=-90] +(0.5,0.45);
\draw (c) to[out=165,in=-90] +(-0.4,0.45);
\draw (c) to +(0,-0.4);
\draw (c2) to[out=15,in=-90] +(0.4,0.45);
\draw (c2) to[out=165,in=-90] +(-0.5,0.45);
\draw (c2) to +(0,-0.4);
\node at (-0.2,-0.3) {$X$};
\node at (0.6,-0.3) {$Y$};
\end{tikzpicture}}}
\qquad\quad
\vcenter{\hbox{%
\begin{tikzpicture}[font=\footnotesize]
\node[copier] (c) at (0,0.4) {};
\draw (c)
to[out=15,in=-90] (0.25,0.7);
\draw (c)
to[out=165,in=-90] (-0.25,0.7);
\draw (c) to (0,0);
\node at (0.2,0.1) {$I$};
\end{tikzpicture}}}
=\;
\vcenter{\hbox{%
\begin{tikzpicture}[font=\footnotesize]
\node at (0.2,-0.05) {};
\draw [gray,dashed] (0,0) rectangle (0.45,0.75);
\end{tikzpicture}}}
\ee
expressed using string diagrams. A \define{classical Markov category} is a classical CD category for which every morphism $X\xrightarrow{f}Y$ is \define{unital}, i.e.\ natural with respect to $\vcenter{\hbox{%
\begin{tikzpicture}[font=\footnotesize]
\node[discarder] (d) at (0,0.15) {};
\draw (d) to (0,-0.25);
\end{tikzpicture}}}$ in the sense that $\;\vcenter{\hbox{%
\begin{tikzpicture}[font=\footnotesize]
\node[arrow box] (c) at (0,0) {$f$};
\node[discarder] (d) at (0,0.45) {};
\draw  (c) to (d);
\draw (c) to (0,-0.45);
\end{tikzpicture}}}
\;=\;
\vcenter{\hbox{%
\begin{tikzpicture}[font=\small]
\node[discarder] (d) at (0,0) {};
\draw (d) to (0,-0.5);
\end{tikzpicture}}}$. 
\vspace{-6mm}
A \define{state} on $X$ is a morphism $I\xrightarrow{p}X$, which is drawn as $\vcenter{\hbox{%
\begin{tikzpicture}[font=\small]
\node[state] (p) at (0,0) {\footnotesize$p$};
\node at (0.15,0.35) {\scriptsize$X$};
\node (X) at (0,0.6) {};
\draw (p) to (X);
\end{tikzpicture}}}$\,.
\ed

\bx
Let $\FinStoch$ be the category whose objects are finite sets and where a morphism $X\xrightarrow{f}Y$ is a stochastic map/conditional probability%
\footnote{The reader will notice that the $\xstoch{\;\;\;}$ notation is not used in this article, unlike in our earlier works~\cite{Pa17,PaRu19,PaBayes,PaRuBayes}. The reason is because we do not need to emphasize the distinction between deterministic maps and stochastic maps in this work.}
 from $X$ to $Y$, which, by definition, assigns to each element $x\in X$ a probability measure $f_{x}$ on $Y$, whose value on $y$ is written as $f_{yx}$. The composite of a composable pair $X\xrightarrow{f}Y\xrightarrow{g}Z$ is defined by the Chapman--Kolmogorov equation $(g\circ f)_{zx}:=\sum_{y\in Y}g_{zy}f_{yx}$. The tensor product is the cartesian product of sets and the product $X\times X'\xrightarrow{f\times f'}Y\times Y'$  of stochastic maps $X\xrightarrow{f}Y$ and $X'\xrightarrow{f'}Y'$, and is given by $(f\times f')_{(y,y')(x,x')}:=f_{yx}f_{y'x'}$. The tensor unit is the single element set, often denoted by $\{\bullet\}$. Functions are special kinds of stochastic maps whose probability measures are $\{0,1\}$-valued. In particular, the maps $\D_{X}$ and $!_{X}$ are the stochastic maps associated to the functions $\D_{X}(x):=(x,x)$ and $!_{X}(x)=\bullet$.  A state on $X$ encodes a probability measure on $X$.
\ex

The conditions in~(\ref{eq:MarkovCatA}) suggest that classical Markov categories cannot be extended to the quantum setting due to the universal no-broadcasting theorem~\cite{BBLW} (see also~\cite[Theorem~5.1]{Ma10}). However, there is a way around these restrictions by working with a larger class of morphisms, adding an even and odd grading for morphisms, 
and substituting the commutativity condition for another closely-related condition~\cite{PaBayes}. 

\bd
A \define{quantum CD category} is a $\Z_{2}$-graded symmetric monoidal category $\mathcal{M}$,%
\footnote{This means that there is a functor $\mM\to\mathbb{B}\Z_{2}$ (where $\mathbb{B}\Z_{2}$ is the one object category whose set of morphisms equals $\Z_{2}=\{0,1\}$ and whose composition is defined by addition modulo $2$ in $\Z_{2}$) and a tensor product is defined for all objects and all morphisms of the \emph{same} degree. Morphisms sent to $0$/$1$ are called \define{even}/\define{odd}. Note that the tensor product of morphisms of different degrees is not defined, but the collection of even morphisms is a symmetric monoidal category.}  
where each object $X$ is equipped with an even discarder, an even copy map, and an odd \define{involution} $*_{X}\equiv \vcenter{\hbox{%
\begin{tikzpicture}[font=\footnotesize]
\node[white dot] (s) at (0,0.4) {};
\draw (s) to (0,0.75);
\draw (s) to (0,0.05);
\node at (0.15,0.15) {\scriptsize $X$};
\end{tikzpicture}}}:X\to X$ satisfying the same conditions as a classical CD category, \emph{except} the last condition in~(\ref{eq:MarkovCatA}), and also satisfying the additional conditions 
\be
\vcenter{\hbox{%
\begin{tikzpicture}[font=\footnotesize]
\node[copier] (c) at (0,0.4) {};
\node[star] (R) at (0.25,0.7) {};
\node[star] (L) at (-0.25,0.7) {};
\draw (c)
to[out=15,in=-90] (R);
\draw (R)
to[out=90,in=-90] (-0.25,1.4);
\draw (c)
to[out=165,in=-90] (L);
\draw (L)
to[out=90,in=-90] (0.25,1.4);
\draw (c) to (0,0.1);
\end{tikzpicture}}}
\quad=\quad
\vcenter{\hbox{%
\begin{tikzpicture}[font=\footnotesize]
\node[copier] (c) at (0,0.3) {};
\node[star] (s) at (0,0) {};
\draw (c)
to[out=15,in=-90] (0.25,0.8);
\draw (c)
to[out=165,in=-90] (-0.25,0.8);
\draw (c) to (s);
\draw (s) to (0,-0.3);
\end{tikzpicture}}}
\qquad\qquad
\vcenter{\hbox{%
\begin{tikzpicture}[font=\small]
\node[star] (s1) at (0,0.15) {};
\node[star] (s2) at (0,0.55) {};
\draw (0,-0.15) to (s1);
\draw (s1) to (s2);
\draw (s2) to (0,0.85);
\end{tikzpicture}}}
\quad=\quad
\vcenter{\hbox{%
\begin{tikzpicture}[font=\footnotesize]
\draw (0,-0.15) to (0,0.85);
\end{tikzpicture}}}
\qquad\qquad
\vcenter{\hbox{%
\begin{tikzpicture}[font=\footnotesize]
\node[star] (s) at (0,0) {};
\draw (s) to +(0,-0.5);
\draw (s) to +(0,0.5);
\node at (0.5,-0.35) {$X\otimes Y$};
\end{tikzpicture}}}
=
\vcenter{\hbox{%
\begin{tikzpicture}[font=\footnotesize]
\node[star] (s) at (0,0) {};
\node[star] (s2) at (0.6,0) {};
\draw (s) to +(0,-0.5);
\draw (s) to +(0,0.5);
\draw (s2) to +(0,-0.5);
\draw (s2) to +(0,0.5);
\node at (0.2,-0.35) {$X$};
\node at (0.8,-0.35) {$Y$};
\end{tikzpicture}}}
\qquad
\quad
\vcenter{\hbox{%
\begin{tikzpicture}[font=\footnotesize]
\node[discarder] (d) at (0,0) {};
\node at (0,0.6) {};
\node[star] (s) at (0,-0.3) {};
\draw (d) to (s);
\draw (s) to (0,-0.6);
\node at (0.25,-0.5) {$X$};
\end{tikzpicture}}}
=\;
\vcenter{\hbox{%
\begin{tikzpicture}[font=\footnotesize]
\node[discarder] (d) at (0,-0.2) {};
\node[star] (s) at (0,0.3) {};
\draw (0,-0.7) to (d);
\draw [gray,dashed] (d) to (s);
\draw [gray,dashed] (s) to (0,0.6);
\node at (0.25,-0.6) {$X$};
\end{tikzpicture}}}
\ee
A \define{quantum Markov category} is a quantum CD category in which every morphism is unital.%
\footnote{Unitality is defined differently for odd morphisms. We exclude the details because we will not need this definition here.} 
A morphism $X\xrightarrow{f}Y$ is said to be \define{$*$-preserving} iff $f\circ*_{X}=*_{Y}\circ f.$ 
\ed

\bx
From now on, all $C^*$-algebras will be assumed unital.
Although the category of finite-dimensional $C^*$-algebras and \emph{positive} unital maps (cf. Notation~\ref{not:prelim}) does not form a quantum Markov category (essentially due to the no-broadcasting theorem), this category naturally embeds into a quantum Markov category, allowing the structure of the ambient quantum Markov category to be utilized~\cite{PaBayes}. 
Let $\fdCAlgUY^{\op}$ be the category whose objects are finite-dimensional $C^*$-algebras (see~\cite[Section~2.3]{Pa17} for a review of $C^*$-algebras within a categorical setting).%
\footnote{Every such finite-dimensional $C^*$-algebra is $^*$-isomorphic to a finite direct sum of (square) matrix algebras~\cite[Theorem~5.5]{Fa01}.}
For example, a matrix algebra will be written as $\mathcal{M}_{n}(\C)$ indicating the $C^*$-algebra of complex $n\times n$ matrices.
 A morphism from $\mA$ to $\mB$ in $\fdCAlgUY^{\op}$ is \emph{either} a linear (even) or \emph{conjugate-linear} (odd) unital map $\mB\xrightarrow{F}\mA$. Notice that the function goes backwards because of the superscript ${}^\op$. The tensor product (over $\C$) is the tensor product of finite-dimensional $C^*$-algebras, so that the unit is $\C$. The tensor product of linear maps is defined in the usual way, while the tensor product of conjugate-linear maps can be defined similarly~\cite[Section~9.2.1]{Uh16}. However, note that it does not make sense to define the tensor product of a linear map with a conjugate-linear one. The $*$ operation is the involution on a $C^*$-algebra, which is conjugate-linear. The copy map $\D_{\mA}$ from $\mA$ to $\mA\otimes\mA$ in $\fdCAlgUY^{\op}$ is the multiplication map $\mA\otimes\mA\xrightarrow{\mu_{\mA}}\mA$ determined on elementary tensors by $A_{1}\otimes A_{2}\mapsto A_{1}A_{2}$. The discard map from $\mA$ to $\C$ in $\fdCAlgUY^{\op}$ is defined to be the unit inclusion map $!_{\mA}:\C\to\mA$ sending $\l\in\C$ to $\l1_{\mA}$. A linear map $\mB\xrightarrow{F}\mA$ is $*$-preserving if and only if it sends self-adjoint elements in $\mB$ to self-adjoint elements in $\mA$. For convenience, we will drop the $\op$ and work directly with the unital maps on the algebras from now on. 
\ex

Although we have introduced the categories $\FinStoch$ and $\fdCAlgUY$, we will be more explicit and work mainly with matrix algebras, linear maps, and positive maps in our main results. The abstract setting will mainly be used in the next two sections to provide the general context. 

\section{Bayes maps, conditionals, and a.e.\ equivalence}
\label{sec:bayescondae}

Here, we review two formulations of Bayes' theorem, which we express categorically. Throughout this section, $\mM$ will denote either a classical or quantum Markov category and $\mC$ will denote some (not necessarily monoidal) subcategory of $\mM$. Furthermore, all morphisms will be even from now on. 

\bd
Given states $I\xrightarrow{p}X$ and $I\xrightarrow{q}Y$, a morphism $X\xrightarrow{f}Y$ in $\mM$ is \define{state-preserving} iff $q=f\circ p$, and one writes $(X,p)\xrightarrow{f}(Y,q)$. A \define{left/right Bayes map for $(X,p)\xrightarrow{f}(Y,q)$} is a morphism $\ov f^{\text{\raisebox{-0.6ex}{$L$}}}$/$\ov f^{\text{\raisebox{-0.6ex}{$R$}}}:Y\to X$ in $\mM$ such that
\be
\label{eq:Bayesequation}
\left.
\vcenter{\hbox{%
\begin{tikzpicture}[font=\small]
\node[state] (q) at (0,0) {$q$};
\node[copier] (copier) at (0,0.3) {};
\node[arrow box] (g) at (-0.5,0.95) {$\ov f^{\text{\raisebox{-0.6ex}{$L$}}}$};
\coordinate (X) at (-0.5,1.5);
\coordinate (Y) at (0.5,1.5);
\draw (q) to (copier);
\draw (copier) to[out=165,in=-90] (g);
\draw (copier) to[out=15,in=-90] (Y);
\draw (g) to (X);
\path[scriptstyle]
node at (-0.7,1.45) {$X$}
node at (0.7,1.45) {$Y$};
\end{tikzpicture}}}
\;\;
=
\;\;
\vcenter{\hbox{%
\begin{tikzpicture}[font=\small]
\node[state] (p) at (0,0) {$p$};
\node[copier] (copier) at (0,0.3) {};
\node[arrow box] (f) at (0.5,0.95) {$f$};
\coordinate (X) at (-0.5,1.5);
\coordinate (Y) at (0.5,1.5);
\draw (p) to (copier);
\draw (copier) to[out=165,in=-90] (X);
\draw (copier) to[out=15,in=-90] (f);
\draw (f) to (Y);
\path[scriptstyle]
node at (-0.7,1.45) {$X$}
node at (0.7,1.45) {$Y$};
\end{tikzpicture}}}
\qquad\middle/\qquad
\vcenter{\hbox{%
\begin{tikzpicture}[font=\small,xscale=-1]
\node[state] (q) at (0,0) {$q$};
\node[copier] (copier) at (0,0.3) {};
\node[arrow box] (g) at (-0.5,0.95) {$\ov f^{\text{\raisebox{-0.6ex}{$R$}}}$};
\coordinate (X) at (-0.5,1.5);
\coordinate (Y) at (0.5,1.5);
\draw (q) to (copier);
\draw (copier) to[out=165,in=-90] (g);
\draw (copier) to[out=15,in=-90] (Y);
\draw (g) to (X);
\path[scriptstyle]
node at (-0.7,1.45) {$X$}
node at (0.7,1.45) {$Y$};
\end{tikzpicture}}}
\;\;
=
\;\;
\vcenter{\hbox{%
\begin{tikzpicture}[font=\small,xscale=-1]
\node[state] (p) at (0,0) {$p$};
\node[copier] (copier) at (0,0.3) {};
\node[arrow box] (f) at (0.5,0.95) {$f$};
\coordinate (X) at (-0.5,1.5);
\coordinate (Y) at (0.5,1.5);
\draw (p) to (copier);
\draw (copier) to[out=165,in=-90] (X);
\draw (copier) to[out=15,in=-90] (f);
\draw (f) to (Y);
\path[scriptstyle]
node at (-0.7,1.45) {$X$}
node at (0.7,1.45) {$Y$};
\end{tikzpicture}}}
\right.
\ee
If all morphisms are in $\mC$, then $\ov f^{\text{\raisebox{-0.6ex}{$L$}}}$/$\ov f^{\text{\raisebox{-0.6ex}{$R$}}}$ are said to be \define{left/right Bayesian inverses of $(X,p)\xrightarrow{f}(Y,q)$} (in $\mC$). 
\ed

Bayes maps are automatically state-preserving. If all morphisms are $*$-preserving, then there is no distinction between left and right concepts (this is always the case in classical Markov categories~\cite{ChJa18,PaBayes}).

\bd
Let $I\xrightarrow{s}X\otimes Y$ be a state and let $p$ and $q$ denote its \define{marginals} $I\xrightarrow{s}X\otimes Y\xrightarrow{\pi_{X}}X$ and $I\xrightarrow{s}X\otimes Y\xrightarrow{\pi_{Y}}Y$, respectively. Here, $\pi_{X}$ and $\pi_{Y}$ are the \define{projections}, which are defined as $\pi_{X}:=\big(X\otimes Y\xrightarrow{\id_{X}\times!_{Y}}X\otimes I\cong X\big)$ and $\pi_{Y}:=\big(X\otimes Y\xrightarrow{!_{X}\otimes\id_{Y}}I\otimes Y\cong Y\big)$. A \define{conditional distribution of $s$ given $Y$ / $X$} (or \define{$Y$ / $X$ conditional} for short) is a morphism $Y\xrightarrow{s|_{Y}}X$ \big/ $X\xrightarrow{s|_{X}}Y$ such that 
\be
\left.
\vcenter{\hbox{%
\begin{tikzpicture}[font=\small]
\node[state] (q) at (0,0) {$q$};
\node[copier] (copier) at (0,0.3) {};
\node[arrow box] (g) at (-0.5,0.95) {$s|_{Y}$};
\coordinate (X) at (-0.5,1.5);
\coordinate (Y) at (0.5,1.5);
\draw (q) to (copier);
\draw (copier) to[out=165,in=-90] (g);
\draw (copier) to[out=15,in=-90] (Y);
\draw (g) to (X);
\path[scriptstyle]
node at (-0.7,1.45) {$X$}
node at (0.7,1.45) {$Y$};
\end{tikzpicture}}}
\quad
=
\quad
\vcenter{\hbox{%
\begin{tikzpicture}[font=\small]
\node[state] (omega) at (0,0) {\;$s$\;};
\coordinate (X) at (-0.25,0.55) {};
\coordinate (Y) at (0.25,0.55) {};
\draw (omega) ++(-0.25, 0) to (X);
\draw (omega) ++(0.25, 0) to (Y);
\path[scriptstyle]
node at (-0.45,0.4) {$X$}
node at (0.45,0.4) {$Y$};
\end{tikzpicture}}}
\qquad
\middle/
\qquad
\vcenter{\hbox{%
\begin{tikzpicture}[font=\small]
\node[state] (omega) at (0,0) {\;$s$\;};
\coordinate (X) at (-0.25,0.55) {};
\coordinate (Y) at (0.25,0.55) {};
\draw (omega) ++(-0.25, 0) to (X);
\draw (omega) ++(0.25, 0) to (Y);
\path[scriptstyle]
node at (-0.45,0.4) {$X$}
node at (0.45,0.4) {$Y$};
\end{tikzpicture}}}
\quad
=
\quad
\vcenter{\hbox{%
\begin{tikzpicture}[font=\small]
\node[state] (p) at (0,0) {$p$};
\node[copier] (copier) at (0,0.3) {};
\node[arrow box] (f) at (0.5,0.95) {$s|_{X}$};
\coordinate (X) at (-0.5,1.5);
\coordinate (Y) at (0.5,1.5);
\draw (p) to (copier);
\draw (copier) to[out=165,in=-90] (X);
\draw (copier) to[out=15,in=-90] (f);
\draw (f) to (Y);
\path[scriptstyle]
node at (-0.7,1.45) {$X$}
node at (0.7,1.45) {$Y$};
\end{tikzpicture}}}
\right.
\;\;.
\ee
\ed

\bd
Let $X$ and $Y$ be objects, let $I\xrightarrow{p}X$ be a state and let $f,g:X\rightarrow Y$ be morphisms. The morphism $f$ is said to be \define{left/right $p$-a.e.\ equivalent to} $g$ iff 
\be
\left.
\vcenter{\hbox{
\begin{tikzpicture}[font=\small]
\node[state] (p) at (0,0) {$p$};
\node[copier] (copier) at (0,0.3) {};
\node[arrow box] (f) at (-0.5,0.95) {$f$};
\coordinate (X) at (0.5,1.5);
\coordinate (Y) at (-0.5,1.5);
\draw (p) to (copier);
\draw (copier) to[out=30,in=-90] (X);
\draw (copier) to[out=165,in=-90] (f);
\draw (f) to (Y);
\path[scriptstyle]
node at (-0.7,1.45) {$Y$}
node at (0.7,1.45) {$X$};
\end{tikzpicture}}}
\quad=\quad
\vcenter{\hbox{
\begin{tikzpicture}[font=\small]
\node[state] (p) at (0,0) {$p$};
\node[copier] (copier) at (0,0.3) {};
\node[arrow box] (g) at (-0.5,0.95) {$g$};
\coordinate (X) at (0.5,1.5);
\coordinate (Y) at (-0.5,1.5);
\draw (p) to (copier);
\draw (copier) to[out=30,in=-90] (X);
\draw (copier) to[out=165,in=-90] (g);
\draw (g) to (Y);
\path[scriptstyle]
node at (-0.7,1.45) {$Y$}
node at (0.7,1.45) {$X$};
\end{tikzpicture}}}
\qquad
\middle/
\qquad
\vcenter{\hbox{
\begin{tikzpicture}[font=\small]
\node[state] (p) at (0,0) {$p$};
\node[copier] (copier) at (0,0.3) {};
\node[arrow box] (f) at (0.5,0.95) {$f$};
\coordinate (X) at (-0.5,1.5);
\coordinate (Y) at (0.5,1.5);
\draw (p) to (copier);
\draw (copier) to[out=150,in=-90] (X);
\draw (copier) to[out=15,in=-90] (f);
\draw (f) to (Y);
\path[scriptstyle]
node at (-0.7,1.45) {$X$}
node at (0.7,1.45) {$Y$};
\end{tikzpicture}}}
\quad=\quad
\vcenter{\hbox{
\begin{tikzpicture}[font=\small]
\node[state] (p) at (0,0) {$p$};
\node[copier] (copier) at (0,0.3) {};
\node[arrow box] (g) at (0.5,0.95) {$g$};
\coordinate (X) at (-0.5,1.5);
\coordinate (Y) at (0.5,1.5);
\draw (p) to (copier);
\draw (copier) to[out=150,in=-90] (X);
\draw (copier) to[out=15,in=-90] (g);
\draw (g) to (Y);
\path[scriptstyle]
node at (-0.7,1.45) {$X$}
node at (0.7,1.45) {$Y$};
\end{tikzpicture}}}
\right.
\quad.
\ee
\ed

All of these definitions are quite similar. Indeed, if $\ov f^{\text{\raisebox{-0.6ex}{$L$}}}$ and $\ov f^{\text{\raisebox{-0.6ex}{$R$}}}$ are left and right Bayes maps for some $(X,p)\xrightarrow{f}(Y,q)$, then they are automatically left and right a.e.\ unique, respectively. Furthermore, the $Y$/$X$ conditionals are also left/right a.e.\ unique. A.e.\ equivalence agrees with the standard measure-theoretic notion~\cite[Proposition~5.3, 5.4]{ChJa18} (as well as the $C^*$-algebraic one~\cite[Definition~2.9]{PaRu19}, \cite[Theorem~5.12]{PaBayes}). With these preliminaries, Bayes' theorem can now be expressed in two different ways. 

\bt
\label{thm:BayesInversion}
[Bayes' theorem via Bayesian inversion]
Every state-preserving stochastic map $(X,p)\xrightarrow{f}(Y,q)$ admits a (necessarily a.e.\ unique) Bayesian inverse (in $\FinStoch$). 
\et

\bt
\label{thm:BayesConditional}
[Bayes' theorem via conditional distributions]
Every joint state $\{\bullet\}\xrightarrow{s}X\times Y$ in $\FinStoch$ admits both (necessarily a.e.\ unique) $X$ and $Y$ conditionals.  
\et

\bx
These two versions of Bayes' theorem are often expressed as the equations
\be
p(x|y)p(y)=p(y|x)p(x)
\qquad\text{ and }\qquad
p(x|y)p(y)=p(x,y)=p(y|x)p(x),
\ee
respectively. Although it seems as though the former is a special case of the latter,%
\footnote{This is especially due to the abusive notation of using $p$ for all mathematical objects.}
 notice that the input data for each definition is different. The first version has input data a morphism $X\xrightarrow{f}Y$ and a state $\{\bullet\}\xrightarrow{p}X$ (the state $q$ on $Y$ is obtained via composition). Meanwhile, the second version has input datum a state $\{\bullet\}\xrightarrow{s}X\times Y$. This distinction may seem pedantic, but it is crucial for generalizing to the non-commutative setting~\cite[Remarks~2.46 and~5.96]{PaRuBayes}.
\ex

\begin{notation}
\label{not:prelim}
If $A$ is a matrix, $A^{\dag}$ denotes its conjugate transpose. 
A matrix $A\in\mM_{m}(\C)$ is \define{positive} iff it is \define{self-adjoint} ($A^{\dag}=A$) and its eigenvalues are non-negative, equivalently $A=C^{\dag}C$ for some $C\in\mM_{m}(\C)$.   
The \define{standard matrix units} of $\mM_{m}(\C)$ will be denoted by $E_{ij}^{(m)}$ with $i,j\in\uline{m}$, where $\uline{m}:=\{1,\dots,m\}$. They satisfy $E_{ij}^{(m)}E_{kl}^{(m)}=\de_{jk}E_{il}^{(m)}$, where $\de_{jk}$ is the Kronecker delta taking value $1$ when $j=k$ and $0$ otherwise. 
A linear map $F:\mM_{n}(\C)\rightarrow\mM_{m}(\C)$ is \define{positive} (\define{completely positive}) iff $F$ ($F\otimes\id_{\mM_{k}(\C)}$) sends positive matrices to positive matrices (for all $k\in\N$).  
In terms of the notation at the beginning of Section~\ref{sec:bayescondae}, $\mM=\fdCAlgUY$ and $\mC=\fdCAlgPU$ is the subcategory consisting of (linear) positive unital maps.
If $F:\mM_{n}(\C)\rightarrow\mM_{m}(\C)$ is linear, then $F^*$ denotes its adjoint with respect to the \define{Hilbert--Schmidt inner product} on matrices, i.e.\ $F^*:\mM_{m}(\C)\to\mM_{n}(\C)$ is the unique linear map satisfying $\tr(F^*(A)B)=\tr(AF(B))$ for all $A\in\mM_{m}(\C)$ and $B\in\mM_{n}(\C)$. An example that appears often is the Hilbert--Schmidt dual of the inclusion $\iota_{\mM_{m}(\C)}:\mM_{m}(\C)\rightarrow\mM_{m}(\C)\otimes\mM_{n}(\C)$ sending $A$ to $A\otimes\mathds{1}_{n}$, and is given by the CPU map $\tr_{\mM_{n}(\C)}$, which is called the \define{partial trace}. Explicitly, $\tr_{\mM_{n}(\C)}$ is determined by its action on simple tensors, namely $\tr_{\mM_{n}(\C)}(A\otimes B)=\tr(B)A$, and it satisfies 
\be
\tr_{\mM_{n}(\C)}\big((A\otimes B)(\mathds{1}_{m}\otimes C)\big)=\tr_{\mM_{n}(\C)}\big((\mathds{1}_{m}\otimes C)(A\otimes B)\big)
\ee
for all inputs $A,B,C$. Finally, if $A\in\mM_{m}(\C)$, then $\Ad_{A}:\mM_{m}(\C)\to\mM_{m}(\C)$ denotes the linear map sending $C\in\mM_{m}(\C)$ to $ACA^{\dag}$. 
\end{notation}

\bx
\label{ex:BayesQM}
Let $\mA:=\mM_{m}(\C)$ and $\mB:=\mM_{n}(\C)$ be two matrix algebras. Let $\omega=\tr(\rho\;\cdot\;)$ and $\xi=\tr(\sigma\;\cdot\;)$ be states on $\mA$ and $\mB$, respectively, with respective density matrices. Let $\mB\xrightarrow{F}\mA$ be a unital linear map. If $\s$ is positive definite (so that the state $\xi$ is faithful), then there are unique left and right Bayes maps for $(\mB,\xi)\xrightarrow{F}(\mA,\omega)$. They are respectively given by 
\be
\ov F^{\text{\raisebox{-0.6ex}{$L$}}}(A):=\s^{-1}F^*(\rho A)
\qquad\text{ and }\qquad
\ov F^{\text{\raisebox{-0.6ex}{$R$}}}(A):=F^*(A\rho)\s^{-1}
\ee
for all $A\in\mA$. If $F$ is $*$-preserving, demanding that these two functions be equal%
\footnote{Note that a.e.\ equivalence now reduces to equality since $\xi$ is faithful.}
 is equivalent to demanding that there is a $*$-preserving Bayes map $\ov F$. In this case, its explicit formula is given by (see~\cite[Corollary~5.32]{PaRuBayes})
\be
\label{eq:PetzLeiferBayes}
\ov F(A)=\sqrt{\s^{-1}}F^*\left(\sqrt{\rho}A\sqrt{\rho}\right)\sqrt{\s^{-1}}.
\ee
Hence, if $F$ is positive unital (PU) or completely positive unital (CPU), then so is $\ov F$. The reader will notice that~(\ref{eq:PetzLeiferBayes}) is the formula for the (dual of the) \define{Petz recovery map}~\cite{AcCe82,Pe84,Pe88,BaKn02,Le06}. However, we will later see that the Petz recovery map is distinct from the Bayes map in general. The difference between the Petz recovery map and the Bayes map is more pronounced in the case that $\s$ is \emph{not} positive definite (so that $\xi$ is not faithful), though the details of this will not be discussed here (but see~\cite{PaRuBayes}). 
\ex

Before using this example, we first need to explain how conditionals can be constructed using Bayes maps more abstractly. Afterwards, we will look at several examples by combining the two results.

\section{Constructing conditionals using Bayes maps}

\bt
\label{thm:conditionalsviaBayes}
Given a joint state $I\xrightarrow{s}X\otimes Y$ with marginals $I\xrightarrow{p}X$ and $I\xrightarrow{q}Y$, let $Y\xrightarrow{L}X\otimes Y$ and $X\xrightarrow{R}X\otimes Y$ be left and right Bayes maps for $(X\times Y,s)\xrightarrow{\pi_{Y}}(Y,q)$ and $(X\times Y,s)\xrightarrow{\pi_{X}}(X,p)$, respectively. Then $s|_{X}:=\big(X\xrightarrow{R}X\otimes Y\xrightarrow{\pi_{Y}}Y\big)$ and $s|_{Y}:=\big(Y\xrightarrow{L}X\otimes Y\xrightarrow{\pi_{X}}X\big)$ are $X$ and $Y$ conditionals of $s$, respectively. 
\et

\bprf
By assumption 
\be
\vcenter{\hbox{%
\begin{tikzpicture}[font=\small]
\node[state] (q) at (0,0) {$q$};
\node[copier] (copier) at (0,0.3) {};
\node[arrow box] (g) at (-0.5,0.95) {\;$L$\;};
\coordinate (X) at (-0.7,1.6);
\coordinate (Y) at (-0.3,1.6);
\coordinate (X2) at (0.5,1.6);
\draw (q) to (copier);
\draw (copier) to[out=165,in=-90] (g);
\draw (copier) to[out=15,in=-90] (X2);
\draw (g.north)++(-0.2,0) to (X);
\draw (g.north)++(0.2,0) to (Y);
\path[scriptstyle]
node at (-0.9,1.55) {$X$}
node at (-0.1,1.55) {$Y$}
node at (0.7,1.55) {$Y$};
\end{tikzpicture}}}
%
\!
=
\;\;
%
\vcenter{\hbox{%
\begin{tikzpicture}[font=\small]
\coordinate (di1) at (-0.75,1.6);
\coordinate (di2) at (0.75,1.6);
\coordinate (d) at (0.35,1.05);
\node[discarder] at (d) {};
\node[state] (s) at (0,0) {\;$s$\;};
\node[copier] (c1) at (-0.25,0.25) {};
\node[copier] (c2) at (0.25,0.25) {};
\draw (s) ++(-0.25,0) to (c1);
\draw (s) ++(0.25,0) to (c2);
\draw (c1) to[out=155,in=-90] (-0.75,0.75)
to (di1);
\draw (c2) to[out=25,in=-90] (di2);
\draw (c1) to[out=25,in=-90] (0.35,0.75) to (d);
\draw (c2) to[out=155,in=-90] (-0.35,0.75) to (-0.35,1.6);
\end{tikzpicture}}}
%
\;
=
\;
\vcenter{\hbox{%
\begin{tikzpicture}[font=\small]
\coordinate (di1) at (-0.75,1.6);
\coordinate (di2) at (0.75,1.6);
\coordinate (d) at (0.35,1.05);
\node[state] (s) at (0,0) {\;$s$\;};
\node[copier] (c2) at (0.25,0.25) {};
\draw (s) ++(-0.25,0) to[out=90,in=-90] (di1);
\draw (s) ++(0.25,0) to (c2);
to (di1);
\draw (c2) to[out=25,in=-90] (di2);
\draw (c2) to[out=155,in=-90] (-0.35,1.6);
\end{tikzpicture}}}
\quad
\text{ and }
\quad
\vcenter{\hbox{%
\begin{tikzpicture}[font=\small,xscale=-1]
\node[state] (q) at (0,0) {$p$};
\node[copier] (copier) at (0,0.3) {};
\node[arrow box] (g) at (-0.5,0.95) {\;$R$\;};
\coordinate (X) at (-0.7,1.6);
\coordinate (Y) at (-0.3,1.6);
\coordinate (X2) at (0.5,1.6);
\draw (q) to (copier);
\draw (copier) to[out=165,in=-90] (g);
\draw (copier) to[out=15,in=-90] (X2);
\draw (g.north)++(-0.2,0) to (X);
\draw (g.north)++(0.2,0) to (Y);
\path[scriptstyle]
node at (-0.9,1.55) {$Y$}
node at (-0.1,1.55) {$X$}
node at (0.7,1.55) {$X$};
\end{tikzpicture}}}
%
\!
=
\;\;
%
\vcenter{\hbox{%
\begin{tikzpicture}[font=\small,xscale=-1]
\coordinate (di1) at (-0.75,1.6);
\coordinate (di2) at (0.75,1.6);
\coordinate (d) at (0.35,1.05);
\node[discarder] at (d) {};
\node[state] (s) at (0,0) {\;$s$\;};
\node[copier] (c1) at (-0.25,0.25) {};
\node[copier] (c2) at (0.25,0.25) {};
\draw (s) ++(-0.25,0) to (c1);
\draw (s) ++(0.25,0) to (c2);
\draw (c1) to[out=155,in=-90] (-0.75,0.75)
to (di1);
\draw (c2) to[out=25,in=-90] (di2);
\draw (c1) to[out=25,in=-90] (0.35,0.75) to (d);
\draw (c2) to[out=155,in=-90] (-0.35,0.75) to (-0.35,1.6);
\end{tikzpicture}}}
%
\;
=
\;
\vcenter{\hbox{%
\begin{tikzpicture}[font=\small,xscale=-1]
\coordinate (di1) at (-0.75,1.6);
\coordinate (di2) at (0.75,1.6);
\coordinate (d) at (0.35,1.05);
\node[state] (s) at (0,0) {\;$s$\;};
\node[copier] (c2) at (0.25,0.25) {};
\draw (s) ++(-0.25,0) to[out=90,in=-90] (di1);
\draw (s) ++(0.25,0) to (c2);
to (di1);
\draw (c2) to[out=25,in=-90] (di2);
\draw (c2) to[out=155,in=-90] (-0.35,1.6);
\end{tikzpicture}}}
\;\;.
\ee
The definitions of $s|_{X}$ and $s|_{Y}$ are drawn as 
\be
\vcenter{\hbox{
\begin{tikzpicture}[font=\small]
\node[arrow box] (p) at (0,0) {$s|_{X}$};
\coordinate (X) at (0,0.8);
\draw (0,-0.8) to (p);
\draw (p) to (X);
\path[scriptstyle]
node at (-0.15,-0.7) {$X$}
node at (-0.15,0.7) {$Y$};
\end{tikzpicture}}}
\;
:=
\;
\vcenter{\hbox{
\begin{tikzpicture}[font=\small]
\node[arrow box] (p) at (0,0) {\;\;$R$\;\;};
\coordinate (X) at (-0.25,0.5);
\coordinate (Y) at (0.25,0.8);
\node[discarder] at (X) {};
\draw (0,-0.8) to (p);
\draw (p.north)++(-0.25,0) to (X);
\draw (p.north)++(0.25,0) to (Y);
\end{tikzpicture}}}
\qquad\text{ and }\qquad
\vcenter{\hbox{
\begin{tikzpicture}[font=\small]
\node[arrow box] (p) at (0,0) {$s|_{Y}$};
\coordinate (X) at (0,0.8);
\draw (0,-0.8) to (p);
\draw (p) to (X);
\path[scriptstyle]
node at (-0.15,-0.7) {$Y$}
node at (-0.15,0.7) {$X$};
\end{tikzpicture}}}
\;
:=
\;
\vcenter{\hbox{
\begin{tikzpicture}[font=\small]
\node[arrow box] (p) at (0,0) {\;\;$L$\;\;};
\coordinate (X) at (-0.25,0.8);
\coordinate (Y) at (0.25,0.5);
\node[discarder] at (Y) {};
\draw (0,-0.8) to (p);
\draw (p.north)++(-0.25,0) to (X);
\draw (p.north)++(0.25,0) to (Y);
\end{tikzpicture}}}
\;\;.
\ee
From this, we immediately obtain
\be
\vcenter{\hbox{%
\begin{tikzpicture}[font=\small]
\node[state] (q) at (0,0) {$q$};
\node[copier] (copier) at (0,0.3) {};
\node[arrow box] (g) at (-0.5,0.95) {$s|_{Y}$};
\coordinate (X) at (-0.5,1.5);
\coordinate (Y) at (0.5,1.5);
\draw (q) to (copier);
\draw (copier) to[out=165,in=-90] (g);
\draw (copier) to[out=15,in=-90] (Y);
\draw (g) to (X);
\path[scriptstyle]
node at (-0.7,1.45) {$X$}
node at (0.7,1.45) {$Y$};
\end{tikzpicture}}}
%
\!\!\!
=
\;\;
%
\vcenter{\hbox{%
\begin{tikzpicture}[font=\small]
\node[state] (q) at (0,0) {$q$};
\node[copier] (copier) at (0,0.3) {};
\node[arrow box] (g) at (-0.5,0.95) {\;$L$\;};
\coordinate (X) at (-0.7,1.6);
\node[discarder] (Y) at (-0.3,1.35) {};
\coordinate (X2) at (0.5,1.6);
\draw (q) to (copier);
\draw (copier) to[out=165,in=-90] (g);
\draw (copier) to[out=15,in=-90] (X2);
\draw (g.north)++(-0.2,0) to (X);
\draw (g.north)++(0.2,0) to (Y);
\end{tikzpicture}}}
%
\;
=
\;
\vcenter{\hbox{%
\begin{tikzpicture}[font=\small]
\coordinate (di1) at (-0.75,1.6);
\coordinate (di2) at (0.75,1.6);
\node[discarder] (d) at (-0.35,1.05) {};
\node[state] (s) at (0,0) {\;$s$\;};
\node[copier] (c2) at (0.25,0.25) {};
\draw (s) ++(-0.25,0) to[out=90,in=-90] (di1);
\draw (s) ++(0.25,0) to (c2);
to (di1);
\draw (c2) to[out=25,in=-90] (di2);
\draw (c2) to[out=155,in=-90] (d);
\end{tikzpicture}}}
%
\;
=
\;
%
\vcenter{\hbox{%
\begin{tikzpicture}[font=\small]
\node[state] (omega) at (0,0) {\;$s$\;};
\coordinate (X) at (-0.25,0.55) {};
\coordinate (Y) at (0.25,0.55) {};
\draw (omega) ++(-0.25, 0) to (X);
\draw (omega) ++(0.25, 0) to (Y);
\path[scriptstyle]
node at (-0.45,0.4) {$X$}
node at (0.45,0.4) {$Y$};
\end{tikzpicture}}}
%
\;
=
\;
%
\vcenter{\hbox{%
\begin{tikzpicture}[font=\small,xscale=-1]
\coordinate (di1) at (-0.75,1.6);
\coordinate (di2) at (0.75,1.6);
\node[discarder] (d) at (-0.35,1.05) {};
\node[state] (s) at (0,0) {\;$s$\;};
\node[copier] (c2) at (0.25,0.25) {};
\draw (s) ++(-0.25,0) to[out=90,in=-90] (di1);
\draw (s) ++(0.25,0) to (c2);
to (di1);
\draw (c2) to[out=25,in=-90] (di2);
\draw (c2) to[out=155,in=-90] (d);
\end{tikzpicture}}}
%
\;
=
\;
%
\vcenter{\hbox{%
\begin{tikzpicture}[font=\small,xscale=-1]
\node[state] (q) at (0,0) {$p$};
\node[copier] (copier) at (0,0.3) {};
\node[arrow box] (g) at (-0.5,0.95) {\;$R$\;};
\coordinate (X) at (-0.7,1.6);
\node[discarder] (Y) at (-0.3,1.35) {};
\coordinate (X2) at (0.5,1.6);
\draw (q) to (copier);
\draw (copier) to[out=165,in=-90] (g);
\draw (copier) to[out=15,in=-90] (X2);
\draw (g.north)++(-0.2,0) to (X);
\draw (g.north)++(0.2,0) to (Y);
\end{tikzpicture}}}
%
\;
=
\!\!
\vcenter{\hbox{%
\begin{tikzpicture}[font=\small]
\node[state] (p) at (0,0) {$p$};
\node[copier] (copier) at (0,0.3) {};
\node[arrow box] (f) at (0.5,0.95) {$s|_{X}$};
\coordinate (X) at (-0.5,1.5);
\coordinate (Y) at (0.5,1.5);
\draw (p) to (copier);
\draw (copier) to[out=165,in=-90] (X);
\draw (copier) to[out=15,in=-90] (f);
\draw (f) to (Y);
\path[scriptstyle]
node at (-0.7,1.45) {$X$}
node at (0.7,1.45) {$Y$};
\end{tikzpicture}}}
\;\;,
\ee
which is the desired conclusion. 
\eprf

This theorem, together with the left/right a.e.\ uniqueness of left/right Bayes maps, is useful because it allows us to write down explicit formulas for conditionals in the quantum setting, at least up to the supports of the states. For the remainder of this work, we will focus on applying this to matrix algebras, rather than arbitrary finite-dimensional $C^*$-algebras. 

\bc
\label{cor:formulaforQMconditionals}
Set $\mA:=\mM_{m}(\C)$ and $\mB:=\mM_{n}(\C)$. Let $\zeta\equiv\tr(\nu\;\cdot\;)$ be a state on $\mA\otimes\mB$ (with density matrix $\nu$) whose marginals on $\mA$ and $\mB$ are given by $\zeta\circ\iota_{\mA}=:\omega\equiv\tr(\rho\;\cdot\;)$ and $\zeta\circ\iota_{\mB}=:\xi\equiv\tr(\sigma\;\cdot\;)$, respectively. Suppose that $\rho$ and $\sigma$ are invertible. Then there are unique conditionals $\mB\xrightarrow{F}\mA$ and $\mA\xrightarrow{G}\mB$ given by 
\be
\label{eq:conditionalformulas}
F(B):=\tr_{\mB}\big((\mathds{1}_{m}\otimes B)\nu\big)\rho^{-1}
\qquad\text{ and }\qquad
G(A):=\s^{-1}\tr_{\mA}\big(\nu(A\otimes\mathds{1}_{n})\big).
\ee
The Hilbert--Schmidt duals of these maps are given by 
\be
\label{eq:HSdualsofcondiotionals}
F^*(A)=\tr_{\mA}\Big(\nu\big((\rho^{-1}A)\otimes\mathds{1}_{n}\big)\Big)
\qquad\text{ and }\qquad
G^*(B)=\tr_{\mB}\Big(\big(\mathds{1}_{m}\otimes (B\s^{-1})\big)\nu\Big).
\ee
\ec

\bprf
The first claim follows from Theorem~\ref{thm:conditionalsviaBayes} and Example~\ref{ex:BayesQM}. For instance, the formula for $F$ is given by $F(B)=\iota_{\mA}^{*}\big((1_{\mA}\otimes B)\nu\big)\rho^{-1}=\tr_{\mB}\big((1_{\mA}\otimes B)\nu\big)\rho^{-1}$. The second claim (\ref{eq:HSdualsofcondiotionals}) follows from the definition of the Hilbert--Schmidt inner product and the cyclic properties of the trace.
\eprf

Are the conditionals $F$ and $G$ in Corollary~\ref{cor:formulaforQMconditionals} positive maps? Let's look at some examples. 

\bx
\label{ex:EPRconditionals}
In the notation of Corollary~\ref{cor:formulaforQMconditionals}, take $m=n=2$ and take $\nu$ to be Bohm's EPR density matrix $\nu:=\frac{1}{2}\left[\begin{smallmatrix}0&0&0&0\\0&1&-1&0\\0&-1&1&0\\0&0&0&0\end{smallmatrix}\right]$ corresponding to the pure state $\frac{1}{\sqrt{2}}\big(|\!\up\>\otimes|\!\dn\>-|\!\dn\>\otimes|\!\up\>\big)
$, where $|\!\up\>$ and $|\!\dn\>$ are just $e_{1}=\left[\begin{smallmatrix}1\\0\end{smallmatrix}\right]$ and $e_{2}=\left[\begin{smallmatrix}0\\1\end{smallmatrix}\right]$ expressed in Dirac notation, \cite{Bo51,EPR}. Then the marginal density matrices $\rho$ and $\sigma$ both equal $\frac{1}{2}\mathds{1}_{2}$. Since this is invertible, the conclusions of Corollary~\ref{cor:formulaforQMconditionals} apply. Hence, 
\be
\label{eq:EPRinferencemapF}
F\!\left(\!\begin{bmatrix}a&b\\c&d\end{bmatrix}\!\right)
\!\!=\!\tr_{\mB}\!\!\left(\!\!\begin{bmatrix}[0.85]a&b&0&0\\c&d&0&0\\0&0&a&b\\0&0&c&d\end{bmatrix}\!\!\!\begin{bmatrix}[0.85]0&0&0&0\\0&1&-1&0\\0&-1&1&0\\0&0&0&0\end{bmatrix}\!\!\right)
\!\!=\!\!\begin{bmatrix}d&-b\\-c&a\end{bmatrix}
\!\!=\!\!\begin{bmatrix}0&1\\-1&0\end{bmatrix}\!\!\begin{bmatrix}a&b\\c&d\end{bmatrix}^{T}\!\!\!\begin{bmatrix}0&-1\\1&0\end{bmatrix}\!,
\ee
which shows that $F$ is PU, but not CPU. The same formula is obtained for $G$. It is worth comparing this expression to the one obtained by using the Petz recovery map instead of the Bayes map. The Petz recovery map $\mathfrak{R}:\mA\otimes\mB\rightarrow\mA$ associated to the inclusion $\iota_{\mA}:\mA\rightarrow\mA\otimes\mB$ and the state $\zeta$ on $\mA\otimes\mB$ is given by 
\be
\mathfrak{R}(A\otimes B)=\sqrt{\rho^{-1}}\tr_{\mB}\left(\sqrt{\nu}(A\otimes B)\sqrt{\nu}\right)\sqrt{\rho^{-1}}=4\tr_{\mB}\left(\nu(A\otimes B)\nu\right), 
\ee
where we have used the fact that $\rho=\frac{1}{2}\mathds{1}_{2}$ and $\nu^2=\nu$ (because $\nu$ is a rank 1 density matrix), so that $\sqrt{\nu}=\nu$. 
Precomposing $\mathfrak{R}$ with the inclusion gives $F':=\mathfrak{R}\circ\iota_{\mB}$, which acts as
\be
F'\left(\begin{bmatrix}a&b\\c&d\end{bmatrix}\right)
=\frac{1}{2}\tr_{\mB}\left(\begin{bmatrix}[0.85]0&0&0&0\\0&a+d&-a-d&0\\0&-a-d&a+d&0\\0&0&0&0\end{bmatrix}\right)
=\frac{1}{2}\begin{bmatrix}a+d&0\\0&a+d\end{bmatrix}
=\frac{\tr\left(\left[\begin{smallmatrix}a&b\\c&d\end{smallmatrix}\right]\right)}{2}\mathds{1}_{2}.
\ee
Notice that the map $F'$, obtained using the Petz recovery map, is actually CPU, unlike the conditional $F$ we obtained in~(\ref{eq:EPRinferencemapF}). 
However, which one of these two maps recovers the standard EPR correlations obtained by a standard quantum-mechanical wave collapse argument? 

Suppose that Alice (represented by $\mA$) obtains new evidence (or has belief) in the form of a state $\varphi=\<\up|\;\cdot\;|\up\>$ (for example, suppose that she set up an apparatus to measure the spin and obtained the result spin up). Then by applying the maps $F$ and $F'$ to these states via pullback,%
\footnote{One could equivalently obtain the Hilbert--Schmidt duals and act on the associated density matrices in the Schr\"odinger picture. We illustrate how this is done in other examples later.}
Alice infers that Bob (represented by $\mB$) would obtain the state on $\mB$ given by 
\be
(\varphi\circ F)\left(\begin{bmatrix}a&b\\c&d\end{bmatrix}\right)=d=\left\<\dn\left|\begin{bmatrix}a&b\\c&d\end{bmatrix}\right|\dn\right\>
\quad\text{ and }\quad
(\varphi\circ F')\left(\begin{bmatrix}a&b\\c&d\end{bmatrix}\right)
=\frac{\tr\left(\left[\begin{smallmatrix}a&b\\c&d\end{smallmatrix}\right]\right)}{2}.
\ee
The first map shows that the spin up state for Alice changes to the spin down state for Bob once the map $F$ is applied. On the other hand, the map $F'$ always gives the totally mixed state for Bob. This indicates that $F$ is a more suitable inference map describing the EPR correlations, since $F'$ loses all the entanglement (more precisely, it is an entanglement breaking channel). Note that analogous conclusions hold if the evidence Alice has is the spin in any direction: $F$ will predict the opposite spin for Bob while $F'$ still predicts the totally mixed state.  
\ex

The fact that $F$, and not $F'$, reproduced the EPR correlations suggests that it has its merits and deserves further study (an alternative derivation of the EPR correlations is done via the formalism of conditional density matrices in~\cite[Section~III.B and Section V.A.3]{LeSp13}, but see footnote~\ref{foot:LS} below). Example~\ref{ex:EPRconditionals} also shows that a joint state can have positive conditionals that are not necessarily CPU. But do conditionals always need to be positive? The next example shows that the answers to this question is no. 

\bx
\label{ex:noconditional}
Set $\mA:=\mM_{2}(\C)$ and $\mB:=\mM_{2}(\C)$. A general pure state in $\C^{2}\otimes\C^{2}\cong\C^{4}$ is of the form 
$
|\Psi\>=c_{\up\up}|\!\up\up\>+c_{\up\dn}|\!\up\dn\>+c_{\dn\up}|\!\dn\up\>+c_{\dn\dn}|\!\dn\dn\>,
$
where $c_{\up\up},c_{\up\dn},c_{\dn\up},c_{\dn\dn}\in\C$ satisfy $|c_{\up\up}|^2+|c_{\up\dn}|^2+|c_{\dn\up}|^2+|c_{\dn\dn}|^2=1$ (here $|\!\up\dn\>=|\!\up\>\otimes|\!\dn\>$ and similarly for the other vectors). Given $p\in(0,1)$, set $q:=1-p$ and let $\nu:=|\Psi\>\<\Psi|$ be the density matrix in $\mA\otimes\mB\cong\mM_{4}(\C)$ associated to the pure state with 
$
c_{\up\up}=\sqrt{\frac{p}{2}},\; 
c_{\up\dn}=\sqrt{\frac{q}{2}},\;
c_{\dn\up}=-\sqrt{\frac{p}{2}},\,$ and $
c_{\dn\dn}=\sqrt{\frac{q}{2}}.
$ 
Then 
\be
\nu=\frac{1}{2}\begin{bmatrix}[0.95]p&\sqrt{pq}&-p&\sqrt{pq}\\\sqrt{pq}&q&-\sqrt{pq}&q\\-p&-\sqrt{pq}&p&-\sqrt{pq}\\\sqrt{pq}&q&-\sqrt{pq}&q\end{bmatrix},
\quad
\rho=\frac{1}{2}\begin{bmatrix}[0.95]1&q-p\\q-p&1\end{bmatrix},
\quad\text{and}\quad
\sigma=\begin{bmatrix}[0.95]p&0\\0&q\end{bmatrix}
.
\ee
Thus, $\rho^{-1}=\frac{1}{2pq}\left[\begin{smallmatrix}1&p-q\\p-q&1\end{smallmatrix}\right]$ and $\sigma^{-1}=\left[\begin{smallmatrix}p^{-1}&0\\0&q^{-1}\end{smallmatrix}\right]$. Using Corollary~\ref{cor:formulaforQMconditionals}, one obtains the explicit formulas 
\be
F\!\left(\left[\begin{smallmatrix}a&b\\c&d\end{smallmatrix}\right]\right)\!=\!\frac{1}{2}\!\begin{bmatrix}a+d+\frac{pc+qb}{\sqrt{pq}}&d-a+\frac{pc-qb}{\sqrt{pq}}\\d-a+\frac{qb-pc}{\sqrt{pq}}&a+d-\frac{pc+qb}{\sqrt{pq}}\end{bmatrix}
,\;\;
G^*\!\left(\left[\begin{smallmatrix}a&b\\c&d\end{smallmatrix}\right]\right)\!=\!\frac{1}{2}\!\begin{bmatrix}a+d+\frac{pb+qc}{\sqrt{pq}}&d-a+\frac{qc-pb}{\sqrt{pq}}\\d-a+\frac{pb-qc}{\sqrt{pq}}&a+d-\frac{pb+qc}{\sqrt{pq}}\end{bmatrix},
\ee
\be
G\left(\left[\begin{smallmatrix}a&b\\c&d\end{smallmatrix}\right]\right)=\frac{1}{2}\begin{bmatrix}a-b-c+d&\sqrt{\frac{q}{p}}(a-b+c-d)\\\sqrt{\frac{p}{q}}(a+b-c-d)&a+b+c+d\end{bmatrix},\quad\text{ and }
\ee
\be
F^*\left(\left[\begin{smallmatrix}a&b\\c&d\end{smallmatrix}\right]\right)=\frac{1}{2}\begin{bmatrix}a-b-c+d&\sqrt{\frac{p}{q}}(a-b+c-d)\\\sqrt{\frac{q}{p}}(a+b-c-d)&a+b+c+d\end{bmatrix}.
\ee
If $p=q=\frac{1}{2}$, then all of these maps are positive. 
Indeed, given a positive matrix of the form $C:=\left[\begin{smallmatrix}a\\b\end{smallmatrix}\right]\left[\begin{smallmatrix}\ov a&\ov b\end{smallmatrix}\right]=\left[\begin{smallmatrix}\ov a a&a\ov b\\b\ov a&b\ov b\end{smallmatrix}\right]$, $F$ and $G$ send this matrix to%
\footnote{All positive matrices are non-negative sums of matrices of this type. Hence, proving $F(C)$ and $G(C)$ are positive is sufficient to proving that $F$ and $G$ are positive, respectively. Proving $F$ and $G$ are positive is also equivalent to proving $F^*$ and $G^*$ are positive.}
\be
F(C)=\frac{1}{2}\begin{bmatrix}\ov a+\ov b\\\ov b-\ov a\end{bmatrix}\begin{bmatrix}a+b&b-a\end{bmatrix}
\;\;\text{ and }\;\;
G(C)=\frac{1}{2}\begin{bmatrix}\ov a-\ov b\\\ov a+\ov b\end{bmatrix}\begin{bmatrix}a-b&a+b\end{bmatrix}\;\;\text{ when $p=q=\frac{1}{2}$}. 
\ee
However, when $p\ne\frac{1}{2}$, then neither $F$ nor $G$ are positive. In fact, neither $F$ nor $G$ are $*$-preserving, which is a necessary condition for positivity. We will come back to this in the next section. 
\ex

\section{Positive conditionals}

Although the conditionals in Corollary~\ref{cor:formulaforQMconditionals} reproduce EPR correlations, the expressions (\ref{eq:conditionalformulas}) and (\ref{eq:HSdualsofcondiotionals}) have two disadvantages. First, they are only partially defined on the supports.%
\footnote{We have not discussed this aspect here because we assumed the marginals are invertible. See~\cite{PaRuBayes} for more details regarding supports and their role in Bayesian inversion. The analogous situation for conditionals is part of ongoing work.}
Second, they need not be positive maps. A necessary condition for positivity is $*$-preservation, so we will first analyze when conditionals are $*$-preserving. 

\blem
\label{lem:necessaryconditional}
Let $\mA:=\mM_{m}(\C),\mB:=\mM_{n}(\C)$, $\zeta=\tr(\nu\;\cdot\;),\omega=\tr(\rho\;\cdot\;),$ $\xi=\tr(\sigma\;\cdot\;)$, $F$, and $G$ be as in Corollary~\ref{cor:formulaforQMconditionals}. Then the following are equivalent. 
\begin{enumerate}[i.]
\itemsep0pt
\item
\label{item:starprescondexists}
A $*$-preserving conditional $\mB\xrightarrow{\zeta|_{\mA}}\mA$ (resp.\ $\mA\xrightarrow{\zeta|_{\mB}}\mB$) exists. 
\item
\label{item:Fdagpreservingconditional}
$\left[\rho,\tr_{\mB}\big(\nu(\mathds{1}_{m}\otimes B)\big)\right]=0$ for all $B\in\mB$ (resp.\ $\left[\sigma,\tr_{\mA}\big(\nu(A\otimes\mathds{1}_{n})\big)\right]=0$ for all $A\in\mA$).
\item
\label{item:Fdagpreservingconditionalmatrixunits}
$\big[\rho,\tr_{\mB}\big(\nu(\mathds{1}_{m}\otimes E_{kl}^{(n)})\big)\big]=0$ for all $k,l\in\uline{n}$ (resp.\ $\big[\sigma,\tr_{\mA}\big(\nu(E_{ij}^{(m)}\otimes\mathds{1}_{n})\big)\big]=0$ for all $i,j\in\uline{m}$).
\item
\label{item:modulargroupcond}
$\Ad_{\rho^{it}}\circ F=F$ (resp.\ $\Ad_{\s^{it}}\circ G=G$) for all $t\in\R$.%
\footnote{$\Ad_{\rho^{it}}$ is the Tomita--Takesaki modular group~\cite{Tak70book} associated to $\omega=\tr(\rho\;\cdot\;)$. This condition is reminiscent of, but distinct from, the Accardi--Cecchini condition~\cite[Proposition~6.1]{AcCe82}.}
\end{enumerate}
\elem

\bprf
(\ref{item:starprescondexists}$\Rightarrow$\ref{item:Fdagpreservingconditional}) 
By Corollary~\ref{cor:formulaforQMconditionals}, the formula for $\zeta|_{\mA}$ is given by $F$. This linear map is $*$-preserving if and only if $F(B^{\dag})=F(B)^{\dag}$, or equivalently $F(B^{\dag})^{\dag}=F(B)$, for all $B$. Assuming $F$ is $*$-preserving, then
\be
\tr_{\mB}\big((\mathds{1}_{m}\otimes B)\nu\big)\rho^{-1}=F(B)=F(B^{\dag})^{\dag}=\left(\tr_{\mB}\Big(\big(\mathds{1}_{m}\otimes B^{\dag}\big)\nu\Big)\rho^{-1}\right)^{\dag}=\rho^{-1}\tr_{\mB}\big(\nu(\mathds{1}_{m}\otimes B)\big).
\ee
Multiplying both sides by $\rho$ and using the properties of the partial trace gives
$
\tr_{\mB}\big(\nu(\mathds{1}_{m}\otimes B)\big)\rho=\rho\tr_{\mB}\big(\nu(\mathds{1}_{m}\otimes B)\big). 
$

\noindent
(\ref{item:Fdagpreservingconditional}$\Rightarrow$\ref{item:starprescondexists}) By a similar calculation, if item~\ref{item:Fdagpreservingconditional} holds, then $F$ is $*$-preserving. 

\noindent
(\ref{item:Fdagpreservingconditional}$\Lra$\ref{item:Fdagpreservingconditionalmatrixunits})
This follows from linearity since every $B$ can be expressed as $B=\sum_{k,l}B_{kl}E_{kl}^{(n)}$. 

\noindent
(\ref{item:modulargroupcond}$\Rightarrow$\ref{item:Fdagpreservingconditional}) By analytic continuation (which is valid due to the finite-dimensionality assumption), $\Ad_{\rho^{it}}\circ F=F$ for all $t\in\C$. Taking $t=i$ in $\Ad_{\rho^{it}}(F(B))=F(B)$ and multiplying both sides by $\rho^2$ reproduces item~\ref{item:Fdagpreservingconditional}. 

\noindent
(\ref{item:Fdagpreservingconditional}$\Rightarrow$\ref{item:modulargroupcond}) By the functional calculus, 
every function of $\rho$ commutes with $\tr_{\mB}\big(\nu(\mathds{1}_{m}\otimes B)\big)$. In particular, $\rho^{z}\tr_{\mB}\big(\nu(\mathds{1}_{m}\otimes B)\big)=\tr_{\mB}\big(\nu(\mathds{1}_{m}\otimes B)\big)\rho^{z}$ for all $z\in\C$. Consequently, $\rho^{it}\tr_{\mB}\big(\nu(\mathds{1}_{m}\otimes B)\big)\rho^{-1}\rho^{-it}=\tr_{\mB}\big(\nu(\mathds{1}_{m}\otimes B)\big)\rho^{-1}$ for all $t\in\R$, which proves item~\ref{item:modulargroupcond}.
\eprf


Lemma~\ref{lem:necessaryconditional} provides a \emph{necessary} condition for \emph{positive} conditionals to exist. Are they sufficient? Namely, if a conditional is $*$-preserving, is it necessarily positive? The motivation for asking this is because the $*$-preserving condition implies (perhaps surprisingly) \emph{complete positivity} for Bayes maps~\cite[Proposition~5.12]{PaRuBayes}. Based on Example~\ref{ex:EPRconditionals}, we so far know that $*$-preservation is \emph{not} strong enough to imply \emph{complete} positivity (or even Schwarz-positivity) for conditionals, so it is natural to ask about positivity alone. In the following theorem, we settle this question in the affirmative. 

\bt
\label{thm:preservingimpliespositiveconditionals}
In the notation of Lemma~\ref{lem:necessaryconditional}, a positive conditional $\mB\xrightarrow{\zeta|_{\mA}}\mA$ (resp.\ $\mA\xrightarrow{\zeta|_{\mB}}\mB$) exists if and only if any (and hence all) of the conditions in Lemma~\ref{lem:necessaryconditional} hold. 
\et

\bprf
It suffices to prove the claim for $F$. 
If $F$ is positive, then it is automatically $*$-preserving, so that one direction in Lemma~\ref{lem:necessaryconditional} applies. Conversely, suppose that $F$ is $*$-preserving. Then by Corollary~\ref{cor:formulaforQMconditionals}, item~\ref{item:Fdagpreservingconditional} of Lemma~\ref{lem:necessaryconditional}, the properties of the partial trace, and the functional calculus%
\footnote{Note that when the assumptions in Theorem~\ref{thm:preservingimpliespositiveconditionals} hold, (\ref{eq:FBBs}) gives an alternative expression for the positive conditional. Namely, $F(B)=\mathrm{Ad}_{\rho^{-1/2}}\left(\tr_{\mB}\left(\nu(\mathds{1}_{m}\otimes B)\right)\right)$. By taking the Hilbert--Schmidt dual, this gives $F^*(A)=\tr_{\mA}\big(\nu\big(\mathrm{Ad}_{\rho^{-1/2}}(A)\otimes\mathds{1}_{n}\big)\big)$, which agrees with the acausal belief propagation map from~\cite[Section~III.B.]{LeSp13}. However, our conditional (cf.\ Corollary~\ref{cor:formulaforQMconditionals}) and the (dual of the) belief propagation map of~\cite{LeSp13} may disagree outside the conditional domain (cf.\ Definition~\ref{defn:commutants}). This is illustrated in Example~\ref{ex:Bobinfersoutsideconditionaldomain}.\label{foot:LS}}
\be
\label{eq:FBBs}
F(B^{\dag}B)=\tr_{\mB}\big((\mathds{1}_{m}\otimes B^{\dag}B)\nu\big)\rho^{-1}
=\sqrt{\rho^{-1}}\tr_{\mB}\big((\mathds{1}_{m}\otimes B)\sqrt{\nu}\sqrt{\nu}(\mathds{1}_{m}\otimes B^{\dag})\big)\sqrt{\rho^{-1}}.
\ee
Since the right-hand-side of this expression is manifestly positive, $F$ is positive. 
\eprf

Even if the $*$-preserving conditions do not hold for \emph{all} elements in the domain algebras, we can always find maximal \emph{subspaces} on which $F$ and $G$ \emph{are} positive. 

\bd
\label{defn:commutants}
In the notation of Lemma~\ref{lem:necessaryconditional}, set
\be
\mA_{\rho^{c}}:=\big\{A\in\mA\;:\;[\rho,A]=0\big\}
\quad\text{ and }\quad
\mB_{\sigma^{c}}:=\big\{B\in\mB\;:\;[\sigma,B]=0\big\}
\ee
to be the \define{commutants} of $\{\rho\}$ and $\{\sigma\}$ inside $\mA$ and $\mB$, respectively. Set
\be
\mB_{\nu}:=\Big\{B\in\mB\;:\;\tr_{\mB}\big(\nu(\mathds{1}_{m}\otimes B)\big)\in\mA_{\rho^{c}}\Big\}
\quad\text{ and }\quad
\mA_{\nu}:=\Big\{A\in\mA\;:\;\tr_{\mA}\big((A\otimes\mathds{1}_{n})\nu\big)\in\mB_{\sigma^{c}}\Big\}
\ee
to be the \define{conditional domains} of $\nu$ inside $\mB$ and $\mA$, respectively. 
A \define{(concrete) operator system} inside $\mM_{k}(\C)$ is a (complex) vector subspace $\mathcal{O}\subseteq\mM_{k}(\C)$ such that $\mathds{1}_{k}\in\mathcal{O}$ and $A\in\mathcal{O}$ implies $A^{\dag}\in\mathcal{O}$. 
\ed

\blem
In the notation of Definition~\ref{defn:commutants}, $\mB_{\nu}$ and $\mA_{\nu}$ are operator systems. 
\elem

\bprf
It suffices to prove this for $\mB_{\nu}$. First, $\mB_{\nu}$ is a subspace by linearity. Second, $\tr_{\mB}(\nu)=\rho$ and $\rho\in\mathcal{A}_{\rho^{c}}$ imply $\mathds{1}_{n}\in\mB_{\nu}$. Third, if $B\in\mB_{\nu}$, then 
\be
\label{eq:Arhocoperatorsystem}
\tr_{\mB}\big(\nu(\mathds{1}_{m}\otimes B^{\dag})\big)=\left(\tr_{\mB}\big((\mathds{1}_{m}\otimes B)\nu\big)\right)^{\dag}=\left(\tr_{\mB}\big(\nu(\mathds{1}_{m}\otimes B)\big)\right)^{\dag}.
\ee
Since $\mA_{\rho^{c}}$ is a $*$-algebra $\tr_{\mB}\big(\nu(\mathds{1}_{m}\otimes B)\big)\in\mA_{\rho^{c}}$ implies $\left(\tr_{\mB}\big(\nu(\mathds{1}_{m}\otimes B)\big)\right)^{\dag}\in\mA_{\rho^{c}}$. Hence, $B^{\dag}\in\mB_{\nu}$ by~(\ref{eq:Arhocoperatorsystem}). Thus, $\mB_{\nu}$ is an operator system. 
\eprf

In this way, although one might not be able to condition on the full algebra to obtain a positive map, one might be able to condition on an operator system inside that algebra. 

\bx
\label{ex:conditionaldomains}
In terms of Example~\ref{ex:noconditional} and assuming $p\ne\frac{1}{2}$, one can show 
\be
\mA_{\rho^{c}}=\left\{\begin{bmatrix}a&b\\b&a\end{bmatrix}\;:\;a,b\in\C\right\}\subset\mA
\quad\text{ and }\quad
\mB_{\sigma^{c}}=\left\{\begin{bmatrix}a&0\\0&d\end{bmatrix}\;:\;a,d\in\C\right\}\subset\mB
\ee
are the commutants. The conditional domains are given by 
\be
\mB_{\nu}=\left\{\begin{bmatrix}a&0\\0&d\end{bmatrix}\;:\;a,d\in\C\right\}\subset\mB
\quad\text{ and }\quad
\mA_{\nu}=\left\{\begin{bmatrix}a&b\\b&a\end{bmatrix}\;:\;a,b\in\C\right\}\subset\mA.
\ee
\ex

Example~\ref{ex:conditionaldomains} suggests that $\mB_{\nu}$ and $\mA_{\nu}$ are not only operator systems, but they might even be $C^*$-subalgebras. Is this always the case? 
%
The answer to this question is no, though the simplest counterexample I could currently find involves a $9\times9$ rank $2$ density matrix with $\mA=\mM_{3}(\C)$ and $\mB=\mM_{3}(\C)$. Its expression is not particularly enlightening, so I have chosen to not record it here.%
\footnote{
Also, I could not find a $4\times4$ density matrix $\nu$ for which the conditional domains are \emph{not} $C^*$-subalgebras, and I suspect that this may always be the case. I hope to resolve this in future work.}

\bc
In the notation of Lemma~\ref{lem:necessaryconditional} and Definition~\ref{defn:commutants}, the restrictions $\mB_{\nu}\hookrightarrow\mB\xrightarrow{F}\mA$ and $\mA_{\nu}\hookrightarrow\mA\xrightarrow{G}\mB$ are positive unital maps 
from operator systems to $C^*$-algebras. 
In terms of the Hilbert--Schmidt duals (the Schr\"odinger picture), the restrictions $\mA_{\rho^{c}}\hookrightarrow\mA\xrightarrow{F^*}\mB$ and $\mB_{\s^{c}}\hookrightarrow\mB\xrightarrow{G^*}\mA$ are positive trace-preserving maps between $C^*$-algebras. 
\ec

Positivity of the Hilbert--Schmidt duals guarantees that density matrices living in the respective commutants always get sent to density matrices under the conditionals. However, inference can still be made even when positivity fails outside these commutants, as the following example illustrates.

\bx
\label{ex:Bobinfersoutsideconditionaldomain}
Set $|\pm\>:= \frac{1}{\sqrt{2}}\left(|\!\up\>\pm|\!\dn\>\right)$.
In terms of Example~\ref{ex:noconditional} (see also Example~\ref{ex:conditionaldomains}), suppose that Alice obtains new evidence in the form of the density matrix $|-\>\<-|\in\mA_{\rho^{c}}$. Using the conditional $F^*$, she infers Bob ($\mB$) to have density matrix $|\!\up\>\<\up\!|$. The same result is obtained by a standard quantum-mechanical wave collapse argument, using the expression
$|\Psi\>=\sqrt{p}|-\!\up\>+\sqrt{q}|+\!\dn\>$,
as well as by the acausal belief propagation map of~\cite[Section~III.B.]{LeSp13}. \emph{However}, if Alice instead obtains new evidence \,${|\!\up\>\<\up\!|}\notin\mA_{\rho^{c}}$, then $F^*\left(|\!\up\>\<\up\!|\right)=\frac{1}{2}\left[\begin{smallmatrix}1&\sqrt{p/q}\\\sqrt{q/p}&1\end{smallmatrix}\right]$, which is not a density matrix unless $p=\frac{1}{2}$. Nevertheless, this matrix is a \emph{non}-orthogonal projection onto the span of the vector $\sqrt{p}|\!\up\>+\sqrt{q}|\!\dn\>$, which is the pure state obtained by a standard quantum-mechanical wave collapse argument (it is orthogonal when $p=\frac{1}{2}$). Meanwhile, the acausal belief propagation map of~\cite{LeSp13} provides the density matrix $|+\>\<+|=\frac{1}{2}\left[\begin{smallmatrix}1&1\\1&1\end{smallmatrix}\right]$ for \emph{all} $p\in(0,1)$, which seems to lose this correlated information. Further investigation is necessary to establish the differences between these approaches.
\ex

\section{Discussion and future directions}

The work presented here investigated conditioning of joint states with faithful marginals for quantum systems from the Markov category perspective. Our definitions are distinct from those of~\cite{Ja18EPTCS}, which defines conditioning in terms of predicates and uses operations analogous to those used to define the Petz recovery map (similar constructions are done using the $Q_{1/2}$ calculus in~\cite{CoSp12,LeSp13}). The root of the distinction between these two approaches comes from our usage of the multiplication map to formulate Bayes maps, even though it is not a positive map. 
By using quantum Markov categories~\cite{PaBayes}, we have been able to define conditioning in a way analogous to what is done in the classical theory, while still using the multiplication map, %
and then finding conditions for which the resulting maps are positive. 

Some work in progress includes the extension of the results presented here to the case where the marginal density matrices are not invertible. Although this seems like an innocent generalization, this is where most of the intricate details occur when analyzing the case of disintegrations and Bayesian inversion in~\cite{PaRu19,PaRuBayes}. It is also what accentuates the difference between Bayesian inverses and the Petz recovery map. Other work in progress includes generalizing Theorem~\ref{thm:preservingimpliespositiveconditionals} for finite-dimensional $C^*$-algebras, i.e.\ direct sums of matrix algebras, to include hybrid classical-quantum systems%
\footnote{The appearance of the modular group in Lemma~\ref{lem:necessaryconditional} and Theorem~\ref{thm:preservingimpliespositiveconditionals} suggests that it may even be possible to extend some of these results to faithful states on certain infinite-dimensional $C^*$-algebras.}
 and to reproduce the conditional version of Bayes' theorem (Theorem~\ref{thm:BayesConditional}) for commutative algebras (Theorem~\ref{thm:BayesInversion} was already reproduced in~\cite[Example~6.33]{PaRuBayes}). 

I see many interesting future directions based on the ideas presented here. For example, what does the set of joint states admitting positive conditionals look like? In what sense can non-positive maps be used to provide inferential information more generally than in Example~\ref{ex:Bobinfersoutsideconditionaldomain}? What is the structure of conditionals for multi-partite (as opposed to bi-partite) states on quantum systems? 
What are the quantum analogues of the theorems describing the consistency of successive conditioning in classical probability theory (cf.\ \cite[Lemma~11.11]{Fr19} and the references therein)? Are there approximate versions of the results presented here using distance measures between states, such as the fidelity or statistical distance? 

\vspace{3mm}
\noindent
{\large \textbf{Acknowledgements.}}
The author thanks the reviewers of QPL'21 for their numerous helpful suggestions.
This research has received funding from the European Research Council (ERC) under the European Union's Horizon 2020 research and innovation program (QUASIFT grant agreement 677368).

\bibliographystyle{eptcs}
\bibliography{conditionalsEPTCS}

\begin{thebibliography}{10}
\providecommand{\bibitemdeclare}[2]{}
\providecommand{\surnamestart}{}
\providecommand{\surnameend}{}
\providecommand{\urlprefix}{Available at }
\providecommand{\url}[1]{\texttt{#1}}
\providecommand{\href}[2]{\texttt{#2}}
\providecommand{\urlalt}[2]{\href{#1}{#2}}
\providecommand{\doi}[1]{doi:\urlalt{http://dx.doi.org/#1}{#1}}
\providecommand{\bibinfo}[2]{#2}

\bibitemdeclare{article}{AcCe82}
\bibitem{AcCe82}
\bibinfo{author}{Luigi \surnamestart Accardi\surnameend} \&
  \bibinfo{author}{Carlo \surnamestart Cecchini\surnameend}
  (\bibinfo{year}{1982}): \emph{\bibinfo{title}{Conditional expectations in von
  {N}eumann algebras and a theorem of {T}akesaki}}.
\newblock {\sl \bibinfo{journal}{J. Funct. Anal.}}
  \bibinfo{volume}{45}(\bibinfo{number}{2}), pp. \bibinfo{pages}{245--273},
  \doi{10.1016/0022-1236(82)90022-2}.

\bibitemdeclare{article}{BBLW}
\bibitem{BBLW}
\bibinfo{author}{Howard \surnamestart Barnum\surnameend},
  \bibinfo{author}{Jonathan \surnamestart Barrett\surnameend},
  \bibinfo{author}{Matthew \surnamestart Leifer\surnameend} \&
  \bibinfo{author}{Alexander \surnamestart Wilce\surnameend}
  (\bibinfo{year}{2007}): \emph{\bibinfo{title}{Generalized No-Broadcasting
  Theorem}}.
\newblock {\sl \bibinfo{journal}{Phys. Rev. Lett.}} \bibinfo{volume}{99}, p.
  \bibinfo{pages}{240501}, \doi{10.1103/PhysRevLett.99.240501}.

\bibitemdeclare{article}{BaKn02}
\bibitem{BaKn02}
\bibinfo{author}{Howard \surnamestart Barnum\surnameend} \&
  \bibinfo{author}{Emanuel \surnamestart Knill\surnameend}
  (\bibinfo{year}{2002}): \emph{\bibinfo{title}{Reversing quantum dynamics with
  near-optimal quantum and classical fidelity}}.
\newblock {\sl \bibinfo{journal}{J. Math. Phys.}}
  \bibinfo{volume}{43}(\bibinfo{number}{5}), pp. \bibinfo{pages}{2097--2106},
  \doi{10.1063/1.1459754}.

\bibitemdeclare{book}{Bo51}
\bibitem{Bo51}
\bibinfo{author}{David \surnamestart Bohm\surnameend} (\bibinfo{year}{1951}):
  \emph{\bibinfo{title}{{Quantum theory}}}.
\newblock \bibinfo{publisher}{Prentice-Hall}, \bibinfo{address}{Englewood
  Cliffs, {NJ}}.
\newblock \bibinfo{note}{Also as reprint ed.: New York, NY, Dover Publications,
  1989}.

\bibitemdeclare{article}{ChJa18}
\bibitem{ChJa18}
\bibinfo{author}{Kenta \surnamestart Cho\surnameend} \& \bibinfo{author}{Bart
  \surnamestart Jacobs\surnameend} (\bibinfo{year}{2019}):
  \emph{\bibinfo{title}{Disintegration and {B}ayesian inversion via string
  diagrams}}.
\newblock {\sl \bibinfo{journal}{Math. Struct. Comp. Sci.}}, pp.
  \bibinfo{pages}{1--34}, \doi{10.1017/S0960129518000488}.

\bibitemdeclare{incollection}{CDDG17}
\bibitem{CDDG17}
\bibinfo{author}{Florence \surnamestart Clerc\surnameend},
  \bibinfo{author}{Vincent \surnamestart Danos\surnameend},
  \bibinfo{author}{Fredrik \surnamestart Dahlqvist\surnameend} \&
  \bibinfo{author}{Ilias \surnamestart Garnier\surnameend}
  (\bibinfo{year}{2017}): \emph{\bibinfo{title}{Pointless learning}}.
\newblock In: {\sl \bibinfo{booktitle}{Foundations of software science and
  computation structures}}, {\sl \bibinfo{series}{Lecture Notes in Comput.
  Sci.}} \bibinfo{volume}{10203}, \bibinfo{publisher}{Springer, Berlin}, pp.
  \bibinfo{pages}{355--369}, \doi{10.1007/978-3-662-54458-7\_21}.

\bibitemdeclare{article}{CoSp12}
\bibitem{CoSp12}
\bibinfo{author}{Bob \surnamestart Coecke\surnameend} \&
  \bibinfo{author}{Robert~W. \surnamestart Spekkens\surnameend}
  (\bibinfo{year}{2012}): \emph{\bibinfo{title}{Picturing classical and quantum
  {B}ayesian inference}}.
\newblock {\sl \bibinfo{journal}{Synthese}}
  \bibinfo{volume}{186}(\bibinfo{number}{3}), pp. \bibinfo{pages}{651--696},
  \doi{10.1007/s11229-011-9917-5}.

\bibitemdeclare{article}{CuSt14}
\bibitem{CuSt14}
\bibinfo{author}{Jared \surnamestart Culbertson\surnameend} \&
  \bibinfo{author}{Kirk \surnamestart Sturtz\surnameend}
  (\bibinfo{year}{2014}): \emph{\bibinfo{title}{A categorical foundation for
  {B}ayesian probability}}.
\newblock {\sl \bibinfo{journal}{Appl. Categ. Structures}}
  \bibinfo{volume}{22}(\bibinfo{number}{4}), pp. \bibinfo{pages}{647--662},
  \doi{10.1007/s10485-013-9324-9}.

\bibitemdeclare{incollection}{DDGK16}
\bibitem{DDGK16}
\bibinfo{author}{Fredrik \surnamestart Dahlqvist\surnameend},
  \bibinfo{author}{Vincent \surnamestart Danos\surnameend},
  \bibinfo{author}{Ilias \surnamestart Garnier\surnameend} \&
  \bibinfo{author}{Ohad \surnamestart Kammar\surnameend}
  (\bibinfo{year}{2016}): \emph{\bibinfo{title}{Bayesian Inversion by
  Omega-Complete Cone Duality}}.
\newblock In \bibinfo{editor}{Jos\'ee \surnamestart Desharnais\surnameend} \&
  \bibinfo{editor}{Radha \surnamestart Jagadeesan\surnameend}, editors: {\sl
  \bibinfo{booktitle}{27th International Conference on Concurrency Theory
  ({CONCUR} 2016)}}, {\sl \bibinfo{series}{Leibniz International Proceedings in
  Informatics ({LIPI}cs)}}~\bibinfo{volume}{59}, \bibinfo{publisher}{Schloss
  Dagstuhl--Leibniz--Zentrum fuer Informatik}, pp. \bibinfo{pages}{1:1--1:15},
  \doi{10.4230/LIPIcs.CONCUR.2016.1}.

\bibitemdeclare{article}{EPR}
\bibitem{EPR}
\bibinfo{author}{Albert \surnamestart Einstein\surnameend},
  \bibinfo{author}{Boris \surnamestart Podolsky\surnameend} \&
  \bibinfo{author}{Nathan \surnamestart Rosen\surnameend}
  (\bibinfo{year}{1935}): \emph{\bibinfo{title}{Can Quantum-Mechanical
  Description of Physical Reality Be Considered Complete?}}
\newblock {\sl \bibinfo{journal}{Phys. Rev.}} \bibinfo{volume}{47}, pp.
  \bibinfo{pages}{777--780}, \doi{10.1103/PhysRev.47.777}.

\bibitemdeclare{book}{Fa01}
\bibitem{Fa01}
\bibinfo{author}{Douglas~R. \surnamestart Farenick\surnameend}
  (\bibinfo{year}{2001}): \emph{\bibinfo{title}{Algebras of linear
  transformations}}.
\newblock \bibinfo{series}{Universitext}, \bibinfo{publisher}{Springer-Verlag,
  New York}, \doi{10.1007/978-1-4613-0097-7}.

\bibitemdeclare{mastersthesis}{Fo12}
\bibitem{Fo12}
\bibinfo{author}{Brendan \surnamestart Fong\surnameend} (\bibinfo{year}{2012}):
  \emph{\bibinfo{title}{Causal theories: A categorical perspective on
  {B}ayesian networks}}.
\newblock Master's thesis, \bibinfo{school}{University of Oxford}.
\newblock \bibinfo{note}{University of Oxford. Available at
  arXiv:\href{https://arxiv.org/abs/1301.6201}{1301.6201 [math.PR]}}.

\bibitemdeclare{article}{Fr19}
\bibitem{Fr19}
\bibinfo{author}{Tobias \surnamestart Fritz\surnameend} (\bibinfo{year}{2020}):
  \emph{\bibinfo{title}{A synthetic approach to {M}arkov kernels, conditional
  independence and theorems on sufficient statistics}}.
\newblock {\sl \bibinfo{journal}{Adv. Math.}} \bibinfo{volume}{370}, p.
  \bibinfo{pages}{107239}, \doi{10.1016/j.aim.2020.107239}.

\bibitemdeclare{article}{FuJa15}
\bibitem{FuJa15}
\bibinfo{author}{Robert \surnamestart Furber\surnameend} \&
  \bibinfo{author}{Bart \surnamestart Jacobs\surnameend}
  (\bibinfo{year}{2015}): \emph{\bibinfo{title}{From {K}leisli categories to
  commutative {$C^*$}-algebras: probabilistic {G}elfand duality}}.
\newblock {\sl \bibinfo{journal}{Log. Methods Comput. Sci.}}
  \bibinfo{volume}{11}(\bibinfo{number}{2}), pp. \bibinfo{pages}{1:5, 28},
  \doi{10.2168/LMCS-11(2:5)2015}.

\bibitemdeclare{inproceedings}{Ja18EPTCS}
\bibitem{Ja18EPTCS}
\bibinfo{author}{Bart \surnamestart Jacobs\surnameend} (\bibinfo{year}{2019}):
  \emph{\bibinfo{title}{Lower and Upper Conditioning in Quantum {B}ayesian
  Theory}}.
\newblock In \bibinfo{editor}{Peter \surnamestart Selinger\surnameend} \&
  \bibinfo{editor}{Giulio \surnamestart Chiribella\surnameend}, editors: {\sl
  \bibinfo{booktitle}{{Proceedings of the 15th International Conference on}
  Quantum Physics and Logic, {{H}alifax, {C}anada, 3-7th June 2018}}}, {\sl
  \bibinfo{series}{Electronic Proceedings in Theoretical Computer Science}}
  \bibinfo{volume}{287}, \bibinfo{publisher}{Open Publishing Association}, pp.
  \bibinfo{pages}{225--238}, \doi{10.4204/EPTCS.287.13}.

\bibitemdeclare{article}{Le06}
\bibitem{Le06}
\bibinfo{author}{Matthew~S. \surnamestart Leifer\surnameend}
  (\bibinfo{year}{2006}): \emph{\bibinfo{title}{Quantum dynamics as an analog
  of conditional probability}}.
\newblock {\sl \bibinfo{journal}{Phys. Rev. A}} \bibinfo{volume}{74}, p.
  \bibinfo{pages}{042310}, \doi{10.1103/PhysRevA.74.042310}.

\bibitemdeclare{article}{LeSp13}
\bibitem{LeSp13}
\bibinfo{author}{Matthew~S. \surnamestart Leifer\surnameend} \&
  \bibinfo{author}{Robert~W. \surnamestart Spekkens\surnameend}
  (\bibinfo{year}{2013}): \emph{\bibinfo{title}{Towards a formulation of
  quantum theory as a causally neutral theory of {B}ayesian inference}}.
\newblock {\sl \bibinfo{journal}{Phys. Rev. A}} \bibinfo{volume}{88}, p.
  \bibinfo{pages}{052130}, \doi{10.1103/PhysRevA.88.052130}.

\bibitemdeclare{incollection}{Ma10}
\bibitem{Ma10}
\bibinfo{author}{Hans \surnamestart Maassen\surnameend} (\bibinfo{year}{2010}):
  \emph{\bibinfo{title}{Quantum probability and quantum information theory}}.
\newblock In: {\sl \bibinfo{booktitle}{Quantum information, computation and
  cryptography}}, \bibinfo{series}{Lect. Notes Physics},
  \bibinfo{publisher}{Springer}, pp. \bibinfo{pages}{65--108},
  \doi{10.1007/978-3-642-11914-9\_3}.

\bibitemdeclare{misc}{Pa17}
\bibitem{Pa17}
\bibinfo{author}{Arthur~J. \surnamestart Parzygnat\surnameend}
  (\bibinfo{year}{2017}): \emph{\bibinfo{title}{Discrete probabilistic and
  algebraic dynamics: a stochastic {G}elfand--{N}aimark Theorem}}.
\newblock \bibinfo{note}{ArXiv preprint:
  \href{https://arxiv.org/abs/1708.00091}{1708.00091 [math.FA]}}.

\bibitemdeclare{misc}{PaBayes}
\bibitem{PaBayes}
\bibinfo{author}{Arthur~J. \surnamestart Parzygnat\surnameend}
  (\bibinfo{year}{2020}): \emph{\bibinfo{title}{Inverses, disintegrations, and
  {B}ayesian inversion in quantum {M}arkov categories}}.
\newblock \bibinfo{note}{ArXiv preprint:
  \href{https://arxiv.org/abs/2001.08375}{2001.08375 [quant-ph]}}.

\bibitemdeclare{misc}{PaRu19}
\bibitem{PaRu19}
\bibinfo{author}{Arthur~J. \surnamestart Parzygnat\surnameend} \&
  \bibinfo{author}{Benjamin~P. \surnamestart Russo\surnameend}
  (\bibinfo{year}{2019}): \emph{\bibinfo{title}{Non-commutative
  disintegrations: existence and uniqueness in finite dimensions}}.
\newblock \bibinfo{note}{ArXiv preprint:
  \href{https://arxiv.org/abs/1907.09689}{1907.09689 [quant-ph]}}.

\bibitemdeclare{misc}{PaRuBayes}
\bibitem{PaRuBayes}
\bibinfo{author}{Arthur~J. \surnamestart Parzygnat\surnameend} \&
  \bibinfo{author}{Benjamin~P. \surnamestart Russo\surnameend}
  (\bibinfo{year}{2020}): \emph{\bibinfo{title}{A non-commutative {B}ayes'
  theorem}}.
\newblock \bibinfo{note}{ArXiv preprint:
  \href{https://arxiv.org/abs/2005.03886}{2005.03886 [quant-ph]}}.

\bibitemdeclare{article}{Pe84}
\bibitem{Pe84}
\bibinfo{author}{D{\'e}nes \surnamestart Petz\surnameend}
  (\bibinfo{year}{1984}): \emph{\bibinfo{title}{A dual in von~{N}eumann
  algebras with weights}}.
\newblock {\sl \bibinfo{journal}{Q. J. Math}}
  \bibinfo{volume}{35}(\bibinfo{number}{4}), pp. \bibinfo{pages}{475--483},
  \doi{10.1093/qmath/35.4.475}.

\bibitemdeclare{article}{Pe88}
\bibitem{Pe88}
\bibinfo{author}{D{\'e}nes \surnamestart Petz\surnameend}
  (\bibinfo{year}{1988}): \emph{\bibinfo{title}{Sufficiency of channels over
  von~{N}eumann algebras}}.
\newblock {\sl \bibinfo{journal}{Q. J. Math.}}
  \bibinfo{volume}{39}(\bibinfo{number}{1}), pp. \bibinfo{pages}{97--108},
  \doi{10.1093/qmath/39.1.97}.

\bibitemdeclare{article}{Se10}
\bibitem{Se10}
\bibinfo{author}{Peter \surnamestart Selinger\surnameend}
  (\bibinfo{year}{2010}): \emph{\bibinfo{title}{A Survey of Graphical Languages
  for Monoidal Categories}}.
\newblock {\sl \bibinfo{journal}{Lect. Notes Phys.}}, pp.
  \bibinfo{pages}{289--355}, \doi{10.1007/978-3-642-12821-9\_4}.

\bibitemdeclare{book}{Tak70book}
\bibitem{Tak70book}
\bibinfo{author}{Masamichi \surnamestart Takesaki\surnameend}
  (\bibinfo{year}{1970}): \emph{\bibinfo{title}{{T}omita's theory of modular
  {H}ilbert algebras and its applications}}.
\newblock {\sl \bibinfo{series}{Lecture Notes in Mathematics}}
  \bibinfo{volume}{128}, \bibinfo{publisher}{Springer},
  \doi{10.1007/BFb0065832}.

\bibitemdeclare{article}{Uh16}
\bibitem{Uh16}
\bibinfo{author}{Armin \surnamestart Uhlmann\surnameend}
  (\bibinfo{year}{2016}): \emph{\bibinfo{title}{Anti- (Conjugate) Linearity}}.
\newblock {\sl \bibinfo{journal}{Sci. China Phys. Mech. Astron.}}
  \bibinfo{volume}{59}(\bibinfo{number}{3}), p. \bibinfo{pages}{630301},
  \doi{10.1007/s11433-015-5777-1}.

\end{thebibliography}
\end{document}